\documentclass[conference]{IEEEtran}
\IEEEoverridecommandlockouts

\usepackage{cite}
\usepackage{amsmath,amssymb,amsfonts}
\usepackage{xspace}
\usepackage{graphicx}
\usepackage{textcomp}
\usepackage{xcolor}
\usepackage{algorithm}
\usepackage{algpseudocode}
\usepackage{tabularx}
\usepackage{svg}
\usepackage{subcaption}
\usepackage{multirow}
\usepackage{booktabs}   
\usepackage{array}     
\usepackage{stfloats}          
\usepackage{booktabs}
\usepackage{float}
\usepackage[section]{placeins}  
\usepackage{amsthm}

\newtheorem{proposition}{Proposition}

\usepackage{adjustbox}
\usepackage{comment}

\def\BibTeX{{\rm B\kern-.05em{\sc i\kern-.025em b}\kern-.08em
    T\kern-.1667em\lower.7ex\hbox{E}\kern-.125emX}}

\newcommand{\name}{RapidGNN\xspace}
\newcommand{\names}{RapidGNN}

\begin{document}

\title{RapidGNN: Communication Efficient Large-Scale Distributed Training of Graph Neural Networks \\
}

\author{\IEEEauthorblockN{Arefin Niam, M S Q Zulkar Nine}
\IEEEauthorblockA{\textit{Department of Computer Science} \\
\textit{Tennessee Technological University}\\
Cookeville, TN, USA \\
aniam42@tntech.edu, mnine@tntech.edu}
}

\maketitle

\begin{abstract}
Graph Neural Networks (GNNs) have achieved state-of-the-art (SOTA) performance in diverse domains. 
However, training GNNs on large-scale graphs poses significant challenges due to high memory demands and significant communication overhead in distributed settings. Traditional sampling-based approaches mitigate computation load to some extent but often fail to address communication inefficiencies inherent in distributed environments. This paper presents \name that introduces a deterministic sampling strategy to precompute mini-batches.
By leveraging the sampling strategy, \name accurately anticipates feature access patterns, enabling optimal cache construction and timely prefetching of remote features. This reduces the frequency and latency of remote data transfers without compromising the stochastic nature of training. Evaluations on Reddit and OGBN-Products datasets demonstrate that \name achieves significant reductions in training time and remote feature fetches, outperforming existing models in both communication efficiency and throughput. Our findings highlight \name's potential for scalable, high-performance GNN training across large, real-world graph datasets along with improving energy efficiency. Our model improves end-to-end training throughput by 2.10$\times$ on average over SOTA model GraphSAGE-METIS (up to \textbf{2.45$\times$} in some settings), while cutting remote feature fetches by over \textbf{4$\times$}. It also reduces energy consumption up to \textbf{23\%}. 

\end{abstract}


\section{Introduction}

Graph Neural Networks (GNNs) have achieved superior results in a multitude of impactful applications through the learning from data structured in a graph format where the relationship between the entities is used to train the models. Recently, GNNs have brought breakthrough results in many tasks in a wide range of scientific fields (e.g., molecular property prediction and drug discovery~\cite{gilmer2017neural,xiong2021graph,li2022multiphysical}, protein structure prediction~\cite{jumper2021highly,jha2022prediction,reau2023deeprank}, material science and crystal structure prediction~\cite{reiser2022graph, batzner20223}, brain connectivity analysis in neuroscience~\cite{bessadok2022graph,kan2022fbnetgen}, particle physics~\cite{shlomi2020graph}, cybersecurity~\cite{bilot2023graph}). GNNs can learn representations based on not only individual entities' features but also the graph dataset's topological structure, enabling it to capture complex relationships among entities. 
Real-world graph datasets are extremely large, for example, the Facebook friendship graph at the time of study~\cite{backstrom2012four} consisted of 721 million users and 69 billion friendship links. As of the fourth quarter of 2024, the social network Meta has 3.35 billion Daily Active People (DAP) across all of its platforms~\cite{meta2025q4}, which also means an exponential increase (e.g., Trillion edges~\cite{ching2015one}) in links across these entities as well. 

Training GNNs in large graphs encounters several problems (e.g., scalability, communication overhead, and workload imbalance). The full batch training of the graph does not scale with memory limitations, necessitating mini-batch training with neighbor sampling. Instead of using the entire graph in one pass, this technique allows selecting a subset from the graph dataset as target nodes and sample nodes from their neighbors through multiple hops to construct smaller computation graphs or blocks. These smaller blocks are repeatedly sampled and passed through the model to update the parameters of the model. While this can reduce memory and computational overhead, it can also lead to new problems in distributed setup. The distributed training of graphs by partitioning the graph across multiple machines can cause communication overhead by frequently fetching large features from other workers. Cai et al.~\cite{cai2021dgcl} shows that the communication overhead in distributed GNN training can take from 50\% to 90\% of the training time. The primary contributor to the communication overhead in distributed GNN training is feature communication during the aggregation phase~\cite{shao2024distributed}. Another problem is the imbalance of workload. Due to their skewed graph pattern, heterogeneous graph structures make it difficult to evenly distribute the load across workers~\cite{raval2017dynamic} in multiple compute nodes.  

A variety of mini-batch sampling strategies have been proposed in the literature. For instance, GraphSAGE~\cite{hamilton2017inductive} introduced node-wise sampling to limit the number of neighbors aggregated per node. FastGCN~\cite{chen2018fastgcn} and LADIES~\cite{zou2019layer} further refined this idea by employing layer-wise sampling and importance sampling techniques, respectively, to reduce computation while maintaining convergence properties. 
However, at the beginning of each iteration, models use these strategies to create computation graphs that require both local and remote node features.
The training process stalls while samplers communicate with remote machines to extract remote features.

Some works in the literature aim to specifically mitigate the communication overhead through system-level solutions and making sampling decisions to reduce overhead~\cite{jiang2021communication}. Distributed GNN training frameworks like DistDGL~\cite{zheng2020distdgl} caches one-hop halo (ghost) nodes (with node IDs)  to avoid communication overhead while constructing the computation block. However, fetching large features from remote partitions significantly contributes to the bottleneck. The P3 system~\cite{gandhi2021p3} implements pipelining of feature communication with computation to hide latency, DGCL~\cite{cai2021dgcl} optimizes data transfer primitives (based on workload and network conditions). It introduces a communication planner that uses a storage hierarchy to schedule peer-to-peer transfers. These works are complementary as they use sampling decision biasing, scheduling, and dynamic caching to hide or mitigate communication delays. However, there remains room for a strategy that \textit{preemptively} avoids as much as communication possible by using the graph computation block itself. Along with hiding communication through pipelining, it is important to reduce the actual communication to effectively reduce the overhead and speed up the training time without impacting the accuracy of the model.

In this paper, we present \name, a novel distributed GNN training framework that aims to minimize the communication overhead at its source and introduce following innovations:
\vspace{-1mm}
\begin{itemize}
    \item We introduce a deterministic sampling strategy using fixed seeds to generate the complete sequence of mini-batches ahead of the training process, enabling us to create efficient caching and prefetching strategies.
    \item We design a novel two-stage feature caching approach using a deterministically precomputed data access pattern. We then use an efficient vector-fetch operation to cache  \(n_{\rm hot}\) “hot” features in bulk RPC operations. At training time, the majority of the remote nodes can accessed through the cache.
    \item We also model a highly efficient asynchronous prefetcher that runs concurrently with the training iterations to prepare mini-batches for the next iteration. The prefetcher effectively pipelines communication with computation and hides communication latency, thus reducing the overall training time. 
    \vspace{-1mm}
\end{itemize}

By incorporating these innovations \name improves end-to-end training throughput by \(\mathbf{2.10\times}\) on average over SOTA model GraphSAGE-METIS (up to \(\mathbf{2.45\times}\) in some settings), cuts remote feature fetches by over 
\(\mathbf{75\%}\) fewer, and reduces the sampling and data copy time by over \(\mathbf{82\%}\), all while matching baseline final accuracy. 

We also observe an overall reduction of \(\mathbf{22 - 23\%}\) in energy consumption during training.

The rest of the paper is organized as follows: Section II provides background on the distributed GNN training, along with related work. Section III illustrates the design of our proposed GNN learning framework. Section IV presents the implementation details of the \name. The  Section V presents extensive experimental evaluations of \texttt{\name}, and Section VI concludes the paper with insights and future research directions.

\section{Background and Related Work}
\subsection{Graph Neural Networks}
A graph can be represented as $G = (V, E)$, where $V = \{v_1, v_2, \ldots, v_n\}$ is the set of nodes and $E \subseteq V \times V$ represents the set of edges. Each node $v_i$ contains feature vector $x_i \in \mathbb{R}^d$. The complete feature space is denoted by $X \in \mathbb{R}^{n \times d}$. If the graph is labeled, each node $v_i$ has corresponding $y_i$ from a label set $Y$.
In GNN training, the node representation is learned by iterative transformation over aggregated neighboring node features. The computation in a GNN layer can be denoted by:
\begin{equation}
h_v^{(l+1)} = \text{COMB}^{(l)}\left(h_v^{(l)}, 
\text{AGG}^{(l)}\left(\{h_u^{(l)} : u \in \eta(v)\}\right)\right)
\end{equation}

Where the feature vector of node $v$ at layer $l$ is denoted by $h_v^{(l)}$ and the set of its neighbors are $\eta(v)$. The Aggregation function, $\text{AGG}^{(l)}$ gathers information from neighboring nodes. Then the combination function $\text{COMB}^{(l)}$ merges the aggregated features with the features of node $v$.

The definition of these functions is arbitrary and dependent on specific GNN architectures (e.g., the weighted sum for aggregation in GCN~\cite{kipf2016semi}, mean/max pooling with concatenation in GraphSAGE~\cite{hamilton2017inductive}).

\subsection{Mini-Batch Training and Sampling in GNNs}
Full‑batch GNN training~\cite{kipf2016semi} quickly exceeds GPU memory on large graphs because each additional layer multiplies the number of reachable neighbors. Researchers, therefore, switch to \textit{Mini‑batch Sampling}~\cite{hamilton2017inductive}, which builds a much smaller computation graph for every iteration. 

In literature, various mini-batch sampling strategies have been proposed. \textit{Node-wise sampling}, as in GraphSAGE~\cite{hamilton2017inductive}, samples a fixed number of neighbors per node to reduce neighborhood explosion, where for each node \( v \) at layer \( l \), a subset \( \widetilde{\eta}(v) = \text{SAMPLE}(\eta(v), k) \) is selected. While it is efficient, it can introduce variance as we mentioned earlier. VR-GCN~\cite{chen2017stochastic} proposes historical activations as control variates: \( h_v^{(l)} = \widetilde{h}_v^{(l)} + h_{v,\text{hist}}^{(l)} - \widetilde{h}_{v,\text{hist}}^{(l)} \). \textit{Layer-wise sampling}, such as FastGCN~\cite{chen2018fastgcn}, samples nodes independently at each layer via importance sampling, while LADIES~\cite{zou2019layer} improves it by enforcing inter-layer connectivity for ensuring meaningful contribution from the sampled nodes. In contrast, \textit{subgraph sampling} strategies like ClusterGCN~\cite{chiang2019cluster} and GraphSAINT~\cite{zeng2019graphsaint} form mini-batches by extracting entire induced subgraphs. For instance, GraphSAINT uses random walks or edge-based sampling and then normalizes the sample-size and importance weight to ensure unbiased gradient estimation: 

\begin{equation}
\nabla_\theta \mathcal{L}(\theta)
\;\approx\;
\frac{1}{|\widetilde{V}|}
\sum_{v \in \widetilde{V}}
\lambda_v\,\nabla_\theta \mathcal{L}_v(\theta)
\end{equation}

Here, we use \(\theta\) as the model parameters, \(\mathcal{L}(\theta)\) as the overall loss, and \(\mathcal{L}_v(\theta)\) as the per-node loss for node \(v\).  Each mini-batch is formed by sampling an induced subgraph \(\widetilde{V}\) (via random walks or edge-based sampling) and assigning each \(v\in\widetilde{V}\) an importance weight \(\lambda_v\) to counter the bias. 

However, this process can introduce sampling variance as only a subset of neighbors contribute per update and can lead to over-smoothing in deeper layers. As messages propagate through many GNN layers, node feature vectors tend to converge to similar values, effectively washing out local structural differences and hurting downstream discrimination. This issue is addressed by carefully designing the scope and depth of sampling.

\subsection{Feature Fetching in Distributed GNN Training}

 In distributed GNN training, the graph dataset is partitioned across multiple machines, with each machine storing a subset of nodes and their features. During training, mini-batches often require multi-hop neighbor features—many of which reside on remote partitions. These features are retrieved via Remote Procedure Calls (RPCs), often synchronously.

Such synchronous remote fetching introduces a significant communication bottleneck. Message passing cannot proceed until all remote features have arrived, which stalls computation and leads to GPU under-utilization. Empirical findings show up to 80\% of training time may be spent on communication and serialization~\cite{gandhi2021p3}. 

One of the most common ways to address this issue is to use a partitioning algorithm to minimize the number of cuts between the connections among the partitions (edge cuts). METIS \cite{karypis1998fast} is the most widely used partitioning algorithm to minimize edge cuts and balance the edges. It attempts to group the mostly connected nodes (therefore, likely to be in the same mini-batch) together. However, having perfect locality for densely connected graphs is impossible. One way to reduce dependency is to truncate the edges. However, this can alter the performance and accuracy of the model. Frameworks like DistDGL~\cite{zhang2023two} uses the \texttt{DistGraph} abstraction and a distributed key-value store (\texttt{KVStore}). However, feature fetching typically remains \textit{on-demand} that keeps the stall time high. Moreover, existing models fetches same feature many times during iteration and epochs.

\subsection{Related Works}

Sampling methods have been one of the key approaches to optimizing and scaling the GNN training frameworks. As discussed in Section II.B, the primary design objective of the sampling methods is to scale the GNN training. The more advanced sampling algorithms aim to reduce computational overhead but also indirectly reduce communication overhead in distributed training setups (mainly by reducing the subgraph size). Some sampling strategies aim to reduce communication overhead by limiting the number of remote nodes sampled through locality-aware sampling. Jiang et al.~\cite {jiang2021communication} skews the neighbor sampling to prioritize local nodes over remote nodes with careful adjustment and ensures that it does not affect convergence much. However, it still has an impact on overall accuracy, and the sampling probabilities are fixed, so it may not adapt well to various configurations. DGS~\cite{wan2022dgs} also follows a similar method but uses an explanation graph to guide the sampling. However, it requires the construction and maintenance of a separate computation graph online that adds to overheads and is subject to the performance of the explanation module.

With graph data distributed across machines, there is very little these strategies can do to limit the communication bottleneck directly. The primary strategy used in these methods to limit communication is partitioning the data using partitioning algorithms like METIS to minimize edge cuts (used in DistDGL\cite{zheng2020distdgl}) to reduce the dependency on remote partitions. However, limiting the communication between partitions through a partitioning algorithm is an NP-hard problem \cite{bazgan2025dense}. Quantization and compression of feature tensors are also used to reduce communication overhead in some works. Sylvie proposed in \cite{zhang2023boosting} uses one-bit quantization for gradient and features, AdaQP~\cite{wan2023adaptive} stochastically quantizes features, embeddings, and gradients into low-precision integers, and in SC-GNN~\cite{wang2024sc} explanation graph is used to prioritize semantically important features. These methods usually have an accuracy trade-off and are subject to rigorous experimental validation. For system-level optimization of communication overhead, the P3~\cite{gandhi2021p3} system introduces a pipelining system to hide the communication in the computation background. P3 improves the utilization of resources but does not reduce the total data transferred over the network. Dorlylus~\cite{thorpe2021dorylus} is another strategy that offloads GNN training to the CPU and uses asynchronous process management for concurrent executions of the training steps. While using serverless computing for GNN training is innovative, it does not address redundant data transfer over the network.

\subsection{Baseline GraphSAGE Model}
Distributed GNN training frameworks like Deep Graph Library (DGL)\cite{wang2019deep} usually fetch the features needed for an iteration of training by dispatching on-the-fly fetch requests for features of each node, which can result in frequent and redundant RPC calls that can dominate training time \cite{gandhi2021p3}. We aim to reduce the communication overhead by reducing the number of RPC calls by identifying exact data access patterns to cache the most used remote nodes' features and minimizing epoch times by prefetching future batches, essentially pipelining the loading of the features with training. For our implementation, we use distributed GraphSAGE from DGL to learn a large graph by partitioning it over multiple machines and then using mini-batch training to update the model parameters. The graph is divided in \(G_{i}\) partitions using Random Partition method\cite{zheng2020distdgl} or METIS\cite{karypis1998fast}.  Each partition is assigned to a training machine and is used by that machine as its local graph partition for running the training process of the GNN and updating the model parameters. The number of training workers and partition should be the same.

After partitioning the partitioned dataset is referenced to the training workers so that each can load their assigned partition. The training device can be both CPU or GPU. 

\begin{figure}[htbp]
    \centering
    \includegraphics[width=0.49\textwidth]{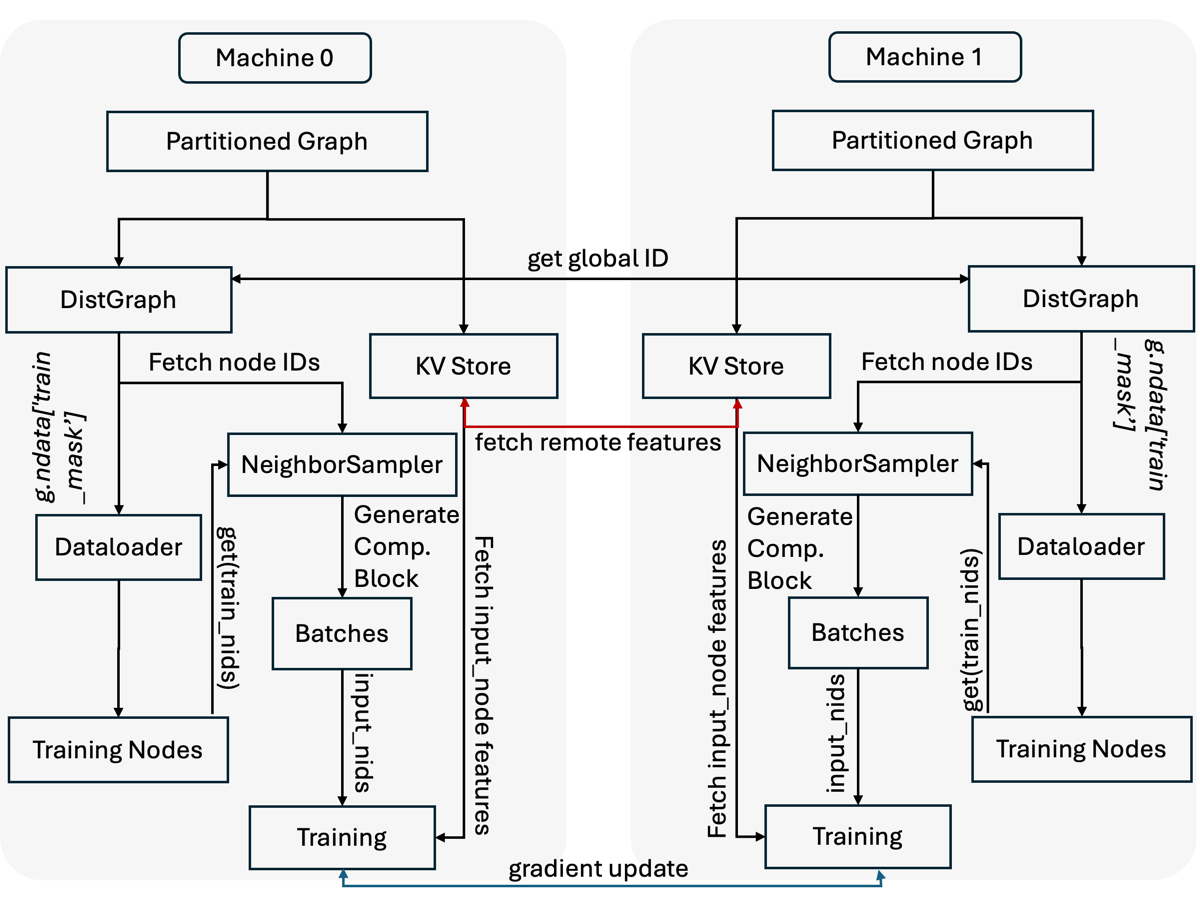} 
    \caption{The working mechanism of baseline GraphSAGE~\cite{hamilton2017inductive} distributed training.}
    \label{fig:graphsage}
    \vspace{-4mm}
\end{figure}


The working mechanism of baseline GraphSAGE distributed training is detailed in Figure \ref{fig:graphsage}. In this example, we take two compute nodes for simplicity of execution and explanation. Each machine gets a part of the partitioned graph dataset and stores them in memory. From the partitioned graph, two objects are obtained: the DistGraph and the KV Store. 

DistGraph provides an abstraction of the graph partitioning so that the local processes can access the whole graph structure when needed using neighborhood sampler. Mainly, it is used to fetch the neighborhood information of the seed nodes to build the computation blocks. On the other hand, the KV Store stores the features of the local nodes and provides a backend mechanism through which the training process can fetch the features from the remote partition during the feature aggregation phase of the training. This fetching of the remote nodes' features during training is one of the primary bottlenecks in communication efficiency as the size of the features is quite large and can stack up due to frequent and redundant access requests. 

The Reddit dataset provided by DGL comprises $232{,}965$ nodes, each represented by a $602$-dimensional feature vector of type \texttt{float32} \cite{dgl_reddit_dataset}. In our profiling run during experiment we found approximately $15{,}000$ nodes to be on remote partition per batch operation. To estimate the network footprint of the node feature tensor:

\begin{itemize}
  \item \textbf{Node feature tensor size:} $232{,}965 \times 602 \times 4\text{ B} \approx 534.7\text{ MB}$.
  
  \item \textbf{Per‐batch transfer} (batch size = 1\,000, 2‐partition setup, $\approx$15\,000 remote nodes/batch): $15{,}000 \times 602 \times 4\text{ B} = 36{,}120{,}000\text{ B} \approx \mathbf{34.45\text{ MiB}}$.
  
  \item \textbf{Batches per epoch:} $\lceil 153{,}431 / 1\,000\rceil = {154}$ 
  
  \item \textbf{Total data per epoch:} $154 \times 36.12\text{ MB} \approx \mathbf{5.6\text{ GB}}$.
\end{itemize}

This can also increase exponentially when more machines are involved, the dataset is larger and batch size increases over a large number of epochs. This highlights the communication overhead in distributed GNN training, where feature data loading can become a significant bottleneck.

Unlike the methods that have been discussed above which react to communication overhead by partitioning, limiting remote node numbers, and quantization/compression at the cost of accuracy and computation overhead, our novel approach proactively reduces communication volume and redundant data fetching operations by using precomputed feature access patterns. By fixing the seeds, we gain a \textit{priori} which remote node features will be needed, when they will be needed, and how often they will be needed to design the caching of most used remote nodes in bulk operations and reuse them. The prefetching mechanism then pipelines the feeding of the features to the training process for upcoming batches. This transforms the system from being a reactive, on-demand process to a coordinated pipeline, yielding a reduction in the number of RPC calls over the network, substantial speedup in training, and reduced energy consumption with minimal changes to the GNN architectures.

\section{RapidGNN}

We address the latency caused by the remote fetching of features at training time in Distributed GNN training by implementing \name, a novel approach that procomputes the remote nodes' feature access pattern by seed assignment to a random neighbor sampler and uses the precomputed access pattern to preemptively cache the most frequently accessed remote nodes' features without affecting the training time. It effectively reduces the number of RPC feature fetch calls. It also designs a prefetcher to rapidly supply the features to the training task, thus improving communication efficiency, training time, and energy efficiency.

\begin{algorithm}[ht]
\caption{\name Training Procedure}
\label{alg:graphcap}
\begin{algorithmic}[1]
\Statex \textbf{Input:} graph $G$; fan‑out $F$; epochs $\mathcal{E}$; cache size $n_{\mathrm{hot}}$; prefetch window $Q$
\Statex \textbf{Output:} trained parameters $\theta$; per‑epoch time $\{t_e\}$; per‑epoch RPCs $\{rpc_e\}$

\State \textit{Precompute} $\{\mathcal{B}_e\}_{e=1}^{\mathcal{E}}$ with fan‑out $F$
\State $N \gets \bigcup_{e,i} N_i^{e}$;\quad $N_{\mathrm{remote}} \gets N\setminus N_{\mathrm{local}}$
\State $N_{\mathrm{cache}} \gets \textit{TopHot}(N_{\mathrm{remote}},\, n_{\mathrm{hot}},\, \textit{freq})$
\State $C_s \gets \textit{VectorPull}(N_{\mathrm{cache}})$
\For{$e = 1$ \textbf{to} $\mathcal{E}$}
    \State $rpc_e \gets 0$;\quad $t_\text{start} \gets \textit{Clock}()$
    \If{$e < \mathcal{E}$}
        \State \textit{Parallel: build} $C_{\mathrm{sec}}$ from $\mathcal{B}_{e+1}$
    \EndIf
    \State \textit{Parallel: prefetch} next $Q$ batches
    \For{$b_i^{e} \in \mathcal{B}_e$}
        \State \textit{GetFeatureFromCache}($N_i^{e}$)
        \If{\textit{miss}}
            \State \textit{SyncPull}($N_i^{e}$);\quad $rpc_e \gets rpc_e + |N_i^{e}|$
        \EndIf
        \State $\theta \gets \textit{Train}(\theta,\, b_i^{e})$
    \EndFor
    \If{$C_{\mathrm{sec}}$ ready}
        \State $C_s \gets C_{\mathrm{sec}}$
    \EndIf
    \State $t_e \gets \textit{Clock}() - t_\text{start}$
\EndFor
\end{algorithmic}
\end{algorithm}


\name (as discussed in Algorithm~\ref{alg:graphcap}) reduces epoch training time $t_{\mathrm{e}}$ and remote RPCs $rpc_{\mathrm{e}}$ by combining deterministic sampling with two-stage caching and asynchronous prefetching. 
%
Mini-batches $\{\mathcal{B}_e\}_{e=1}^\mathcal{E}$ are precomputed using fan-out $F$, and the complete set of accessed nodes is collected (Line 1)
\begin{equation}
N = \bigcup_{e=1}^\mathcal{E} \bigcup_{i=1}^B N^e_i
\end{equation}

(Line 2-3) Remote nodes are identified as $N_{\mathrm{remote}} = N \setminus N_{\mathrm{local}}$, and the most frequent $n_{\mathrm{hot}}$ nodes form the cache set 
\[
N_{\mathrm{cache}} = \{ n \in N_{\mathrm{remote}} \mid \textit{freq}(n) \text{ ranks top-}n_{\mathrm{hot}} \}.
\]
Their features are bulk-fetched via vectorized RPC into a steady cache $C_s$ (Line 4). The fetching of the features from the cache replaces the default on-the-fly fetching mechanism in DGL. 

During the training phase (Line 5-22), for each epoch, a background thread is concurrently launched to precompute a secondary cache $C_{\mathrm{sec}}$ using the mini-batches for the next epoch, $\mathcal{B}_{e+1}$ (Line 8). In parallel, a prefetcher continuously populates a queue (of size $Q$) with upcoming batch features (Line 10). For each batch $b^e_i$, the training loop waits for the prefetched features corresponding to the input nodes $N^e_i$, resorting to a synchronous pull only when necessary (Line 12-14). The combined features, assembled from the steady cache $C_s$ and any missing entries, are then transferred to the GPU with the corresponding computational blocks, after which the standard forward and backward passes and parameter updates are executed (Line 16). At the end of each epoch, if the secondary cache $C_{\mathrm{sec}}$ has been successfully computed, it is swapped into $C_s$, ensuring that the cache remains adaptive to any changes in the sampling pattern (Line 19). As a result, \name minimizes the waiting time for RPC calls by serving the majority of feature requests from the cache and via asynchronous prefetching, thereby reducing both the epoch training time, $t_{\mathrm{e}}$, and the overall number of redundant RPCs, $rpc_{\mathrm{e}}$, compared with a baseline approach that does not incorporate these techniques. 

\section{Implementation}

\name integrates remote nodes' feature caching and asynchronous prefetching mechanism into a scalable distributed GNN training pipeline. We implement our design to augment the DGL framework for mini-batch distributed GNN training. The core operations of \name can be divided into two phases: (1) an offline precomputation stage that determines the feature access patterns of the training process in advance and (2) the online caching and prefetching mechanism that runs concurrently with the training iterations. They utilize the precomputed feature access pattern to preload remote nodes' features and hide the feature loading time parallel to the training task.

At the core of the overall architecture is the pre-computation stage, where all workers use a globally shared random seed for neighbor sampling that is fixed using \texttt{torch.manual} provided in the pytorch framework. The seed is systematically varied across epochs and batches using the configuration numbers so that they never repeat throughout the training while maintaining reproducibility. We preserve stochasticity across training by aligning seed generation with epoch and batch indices while maintaining consistency across distributed workers. This precomputation is done offline to the training and is later used in the training to guide the caching and prefetching mechanism. We also ensure that our deterministic sampling does not hurt the convergence of the training.

\begin{proposition}
Let $\mathcal{B}_e$ be the mini‑batch produced by running a uniform neighbor sampler on graph $G$ with fan‑out $F$ using a pseudorandom generator seeded by
\[
s_e := s_0 + e,
\]
where $s_0$ is fixed and $e = 1,2,\dots,\mathcal{E}$. Assume the PRNG behaves as an ideal uniform source of randomness. Then:
\begin{enumerate}
  \item[(a)] Each $\mathcal{B}_e$ has exactly the same marginal distribution as a truly random mini‑batch of fan‑out~$F$.
  \item[(b)] For any $e\neq e'$, the draws $\mathcal{B}_e$ and $\mathcal{B}_{e'}$ are independent.
  \item[(c)] The stochastic gradient
  \begin{equation}
    g(\theta;\mathcal{B}_e)
    \;=\;
    \nabla_\theta \frac1{|\mathcal{B}_e|}\sum_{v\in\mathcal{B}_e}\mathcal{L}_v(\theta)
\end{equation}
  remains unbiased, i.e., $\mathbb{E}[\,g(\theta;\mathcal{B}_e)\,]=\nabla_\theta \mathcal{L}(\theta)$, and has strictly positive variance.
\end{enumerate}
\end{proposition}

\begin{proof}
(a) A PRNG seeded by $s_e$ is statistically indistinguishable from true i.i.d.\ uniform bits, so sampling neighbors with it produces exactly the same distribution as on‑the‑fly uniform sampling.

(b) Distinct seeds $s_e\neq s_{e'}$ yield non‑overlapping PRNG streams, hence the bit sequences (and resulting batches) are independent.

(c) Since each $\mathcal{B}_e$ is marginally identical to an ideal random draw, standard mini‑batch SGD theory implies
\begin{equation}
\mathbb{E}\bigl[g(\theta;\mathcal{B}_e)\bigr]
=\nabla_\theta \mathcal{L}(\theta),
\quad
\mathrm{Var}\bigl[g(\theta;\mathcal{B}_e)\bigr]>0.
\end{equation}
Independence across epochs then guarantees the usual convergence properties of SGD remain unaltered. The evaluation section provides evidence for the convergence proof.
\end{proof}

We use this seed to run the precomputation offline. We can generate the complete computation block comprising the list of batches within each epoch and the order of the input nodes within them. We get the complete list of nodes per epoch from the precomputed block and identify the remote nodes using the partition book, which contains information on partition ownership. After aggregating the remote node IDs, we count the frequency of their occurrence, as many remote nodes are sampled more than once across the batches of computation blocks. This strategy is mainly the offline precomputation phase that gives us the information needed to guide the rest of the Caching and Prefetching process.

\begin{figure}[htbp]
    \centering
    \includegraphics[width=0.50\textwidth]{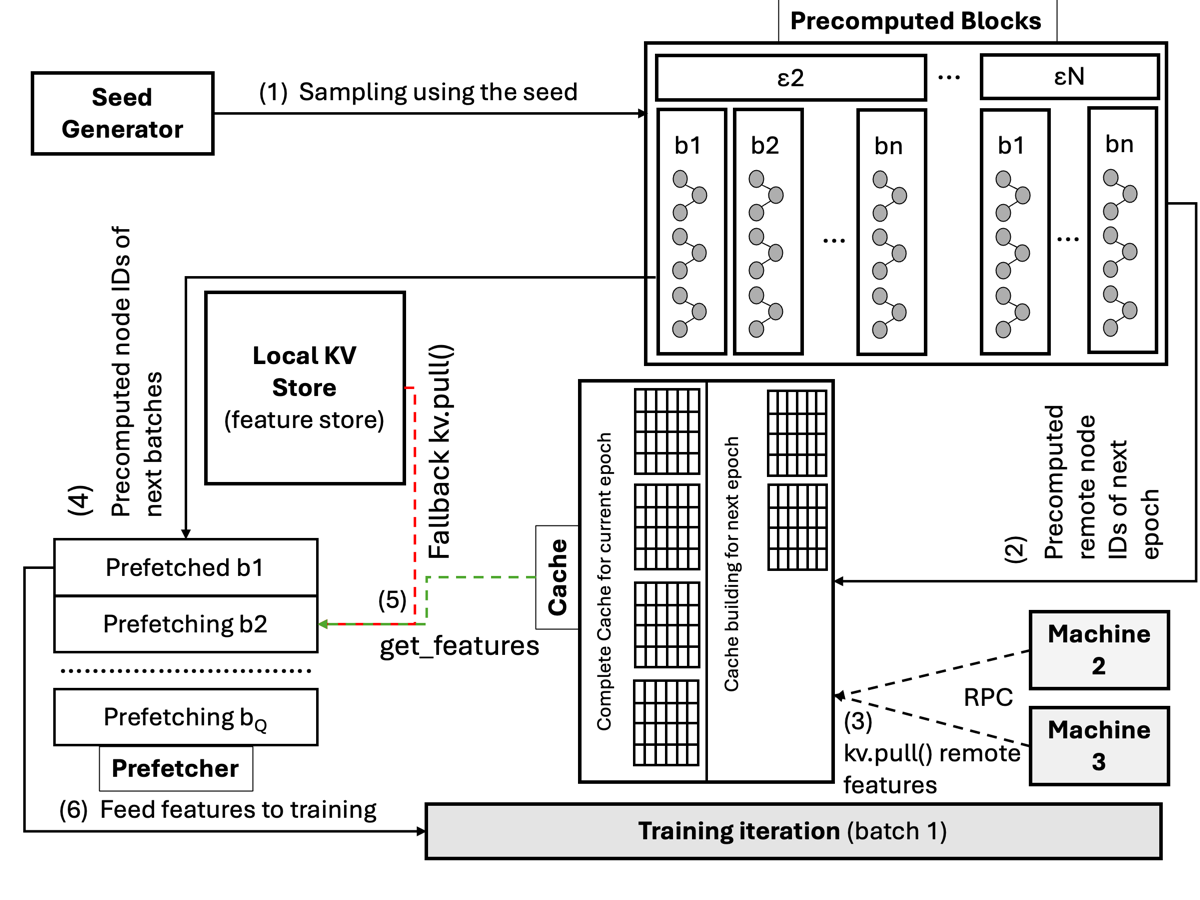} 
    \caption{The working mechanism of \name}
    \label{fig:graphcap}
    \vspace{-2mm}
\end{figure}

In Figure~\ref{fig:graphcap}, we provide a high-level schematic overview of the \name architecture. We show one participating machine in the distributed training as a self-contained unit independently generating a computation block with the sampler using the seed value (Step 1). The double buffer cache uses the generated computation block (Step 2) to cache n-hot remote nodes per epoch (Step 3) from remote machines. The prefetcher uses the precomputed block (Step 4) to fetch the features of the subsequent batches in parallel to the training process. The missing features are retrieved with a fallback mechanism by submitting pull requests to the KV Store (Step 5). The prefetcher prepares the features of future batches, and when requested by the training block, it readily supplies them (Step 6). 

\begin{figure}[htbp]
    \centering
    \includegraphics[width=0.50\textwidth]{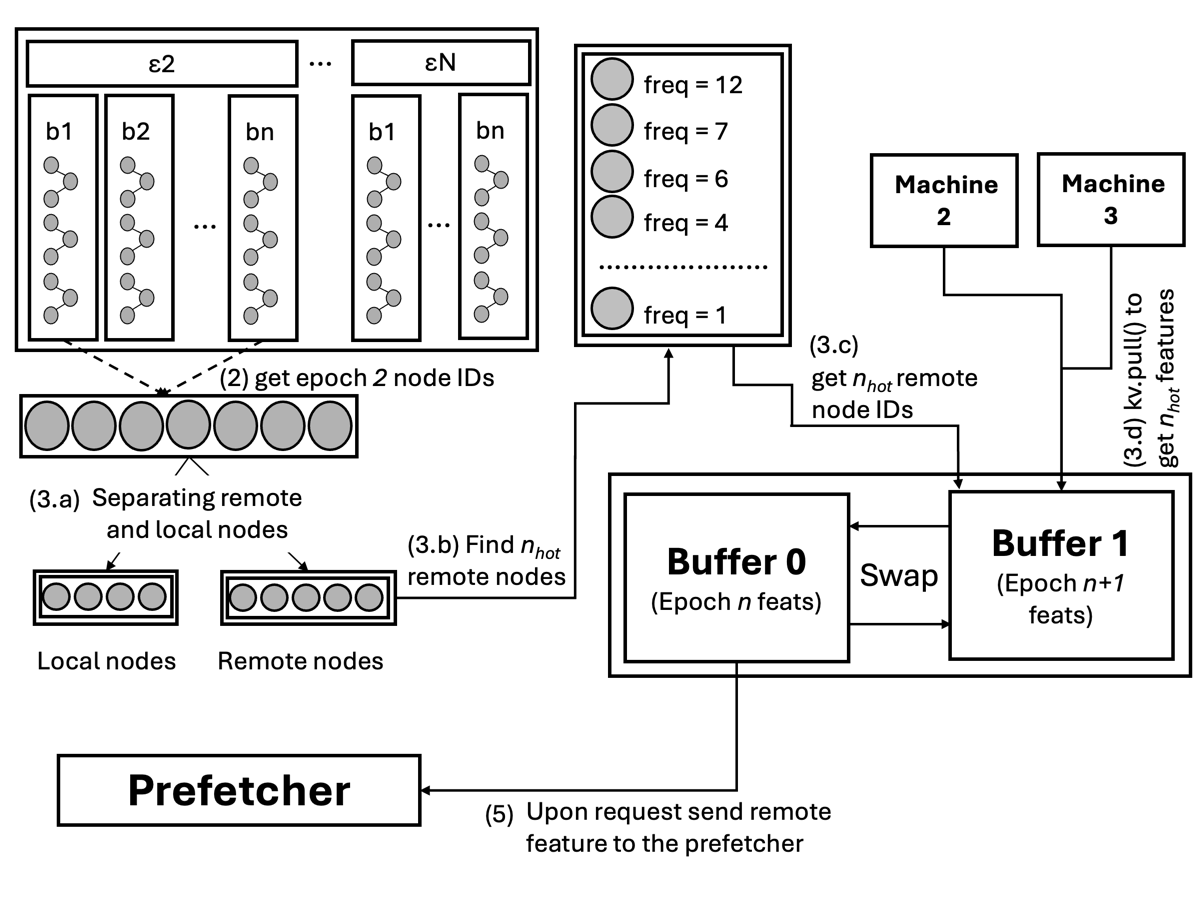} 
    \caption{The working mechanism of \name(Cache Architecture)}
    \label{fig:cache}
\end{figure}

Figure~\ref{fig:cache} demonstrates the cache-building process. To build the cache, we retrieve the node IDs from the computation block (Step 2) and filter out the remote nodes (Step 3.a). Once the access frequencies of the remote nodes are calculated (Step 3.b), the top-\(n_{\mathrm{hot}}\) remote nodes are identified as the primary candidates to be cached (Step 3.c). This selection process is critical in keeping the cache effective; rather than caching nodes based solely on static graph topology, we focus on nodes empirically proven to be accessed frequently across mini-batches.

In the next phase, we build the steady cache \(C_s\) by issuing a single, vectorized RPC call to fetch all the feature vectors of all nodes in \(N_{\mathrm{cache}}\) from the remote KV Store (Step 3.d). This bulk feature fetching operation is drastically more efficient than individually identifying remote nodes and issuing separate calls for each. The fetched features are then stored in the primary slot of the double buffer. When the training begins, the sampler generates the batches for all the epochs (as generated in the precomputation), and we spawn the training process and the prefetching process. 

\begin{figure}[htbp]
    \centering
    \includegraphics[width=0.50\textwidth]{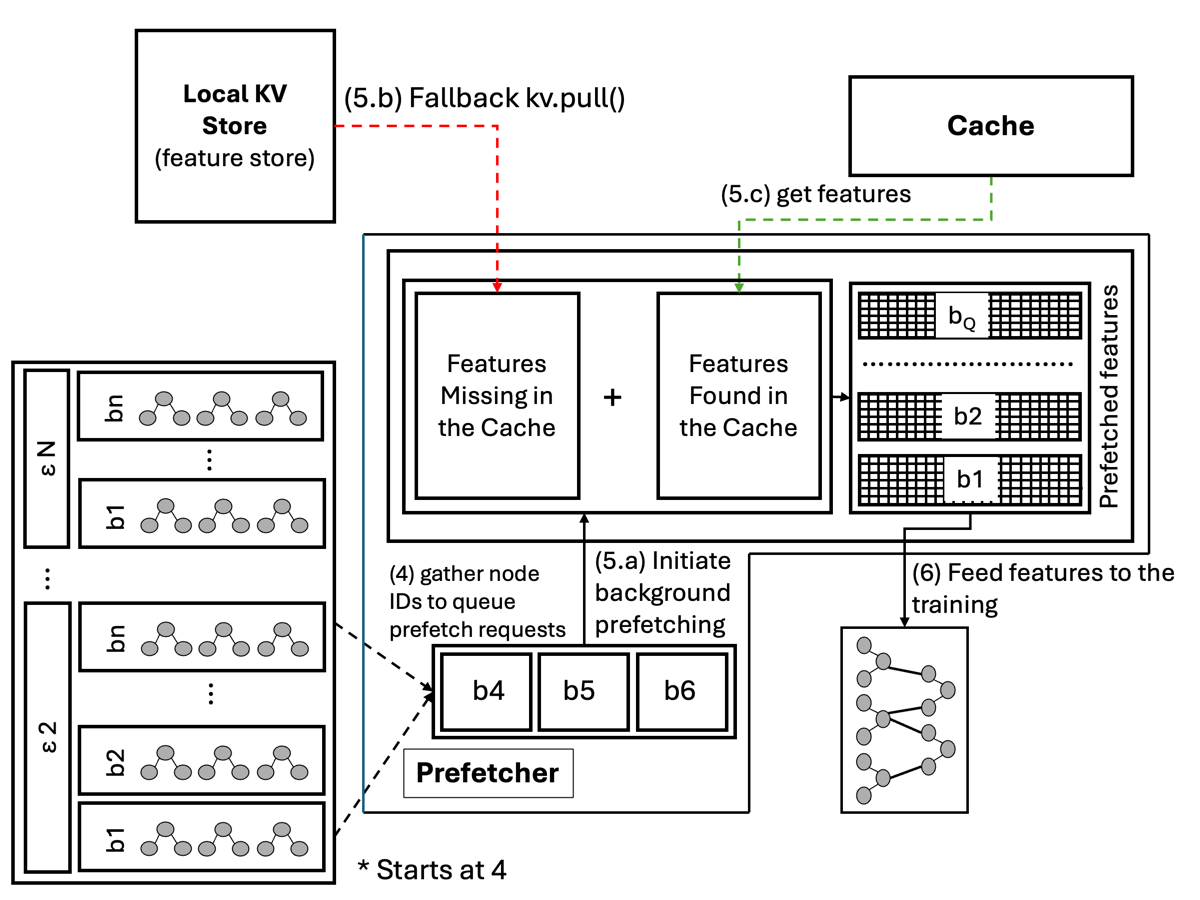} 
    \caption{The working mechanism of \name (Prefetcher)}
    \label{fig:prefetcher}
\end{figure}

In parallel to the training process, we build a steady supply of per batch input node features using the prefetcher shown in Figure~\ref{fig:prefetcher}. The prefetcher fetches the features from the cache  \(C_s\). Any cache misses are fetched through the default KV Store (Step 5.b). We queue the prefetch requests for the upcoming batches in the prefetcher (Step 5.a) and get the features of the immediate next batch (Step 5.c). As we have already stored the most used remote node features in the cache, the number of remote calls is drastically reduced. The training loop instantly accesses the data, and the training proceeds as designed (Step 6).
\section{Evaluation}

To evaluate the effectiveness of \name in reducing training time and communication overhead, we conduct extensive experiments on two benchmark datasets and compare them against the SOTA models. Our evaluation aims to quantify the improvements in training speed, communication reduction, and energy efficiency. We also provide validation of our Proposition 1, showing that the deterministic precomputation does not impact the accuracy of the models. 

\begin{figure*}[!t]
  \centering

  \includegraphics[width=0.6\textwidth]{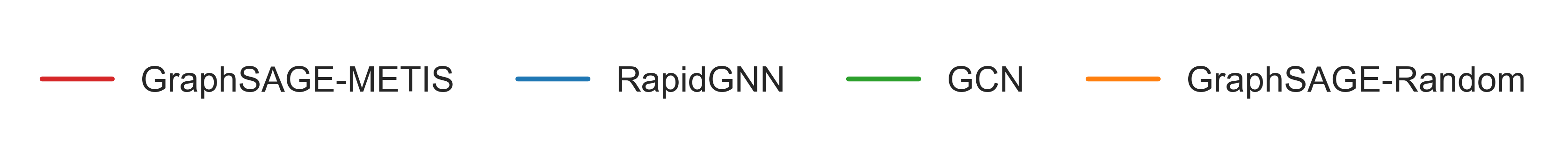}
  \vspace{2mm}

  \begin{subfigure}[b]{0.27\textwidth}
    \centering
    \includegraphics[width=\linewidth]{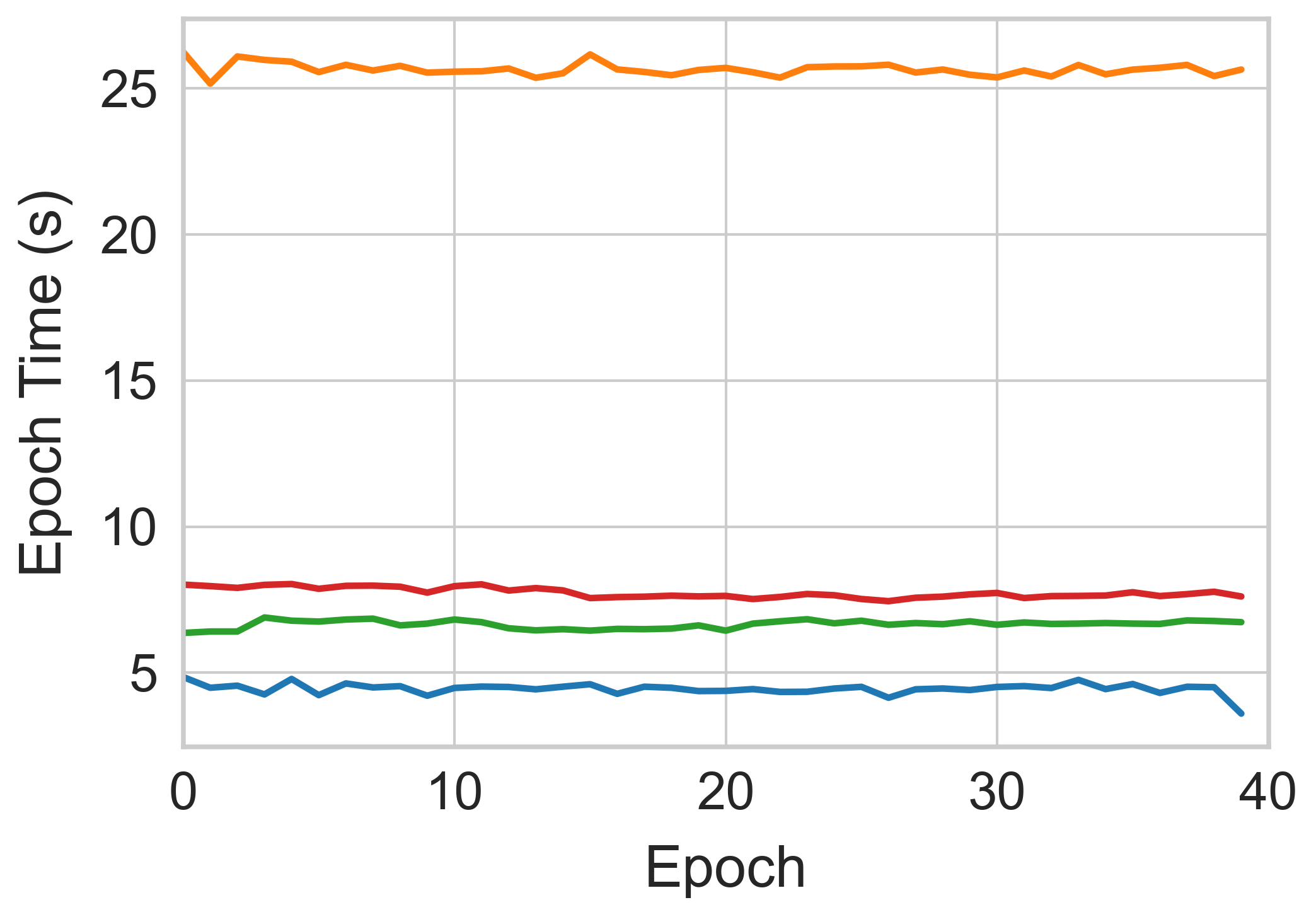}
    \caption{OGBN‑Products, batch size 1000}
    \label{fig:ogbn_1000}
  \end{subfigure}\hfill
  \begin{subfigure}[b]{0.27\textwidth}
    \centering
    \includegraphics[width=\linewidth]{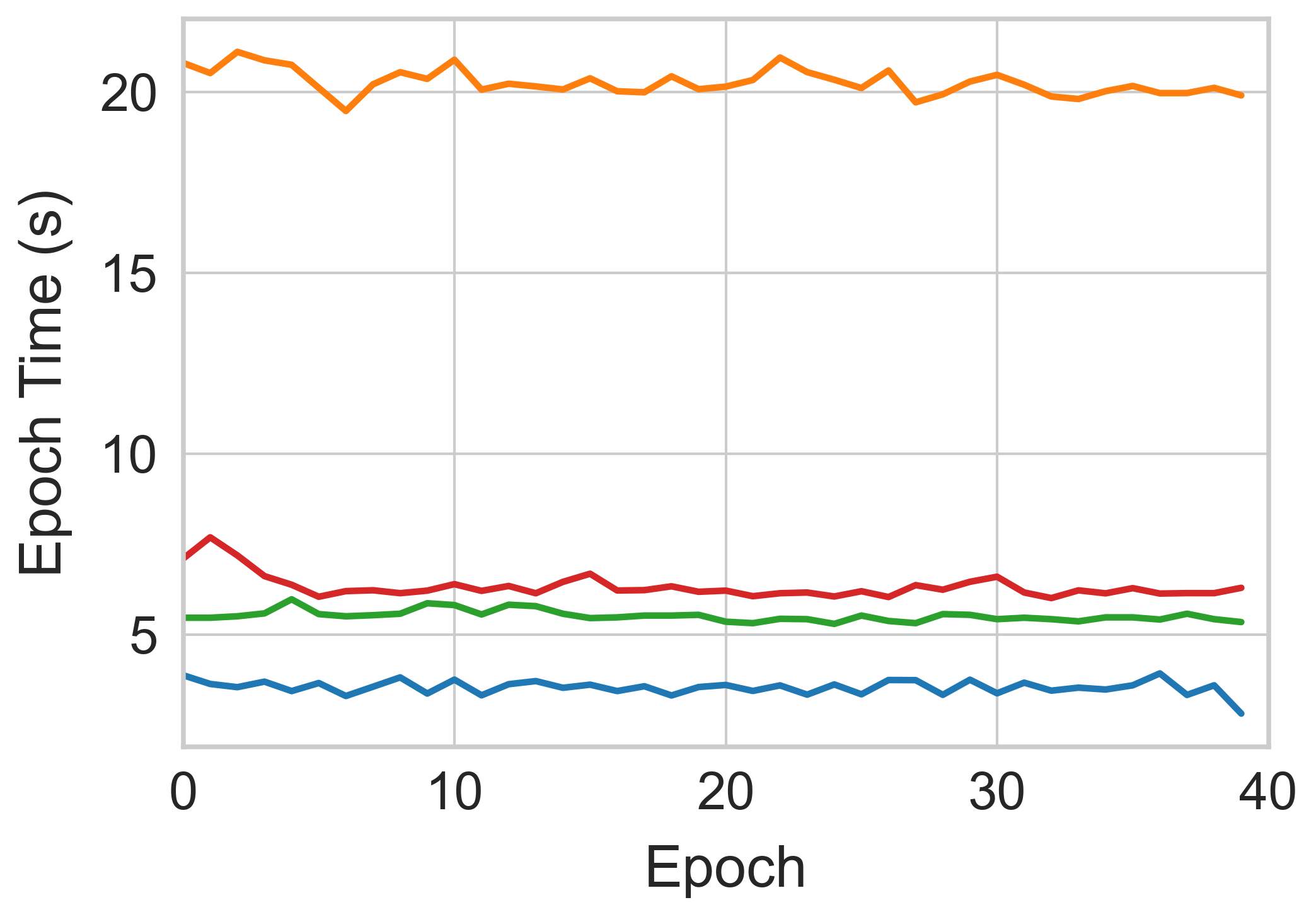}
    \caption{OGBN‑Products, batch size 2000}
    \label{fig:ogbn_2000}
  \end{subfigure}\hfill
  \begin{subfigure}[b]{0.27\textwidth}
    \centering
    \includegraphics[width=\linewidth]{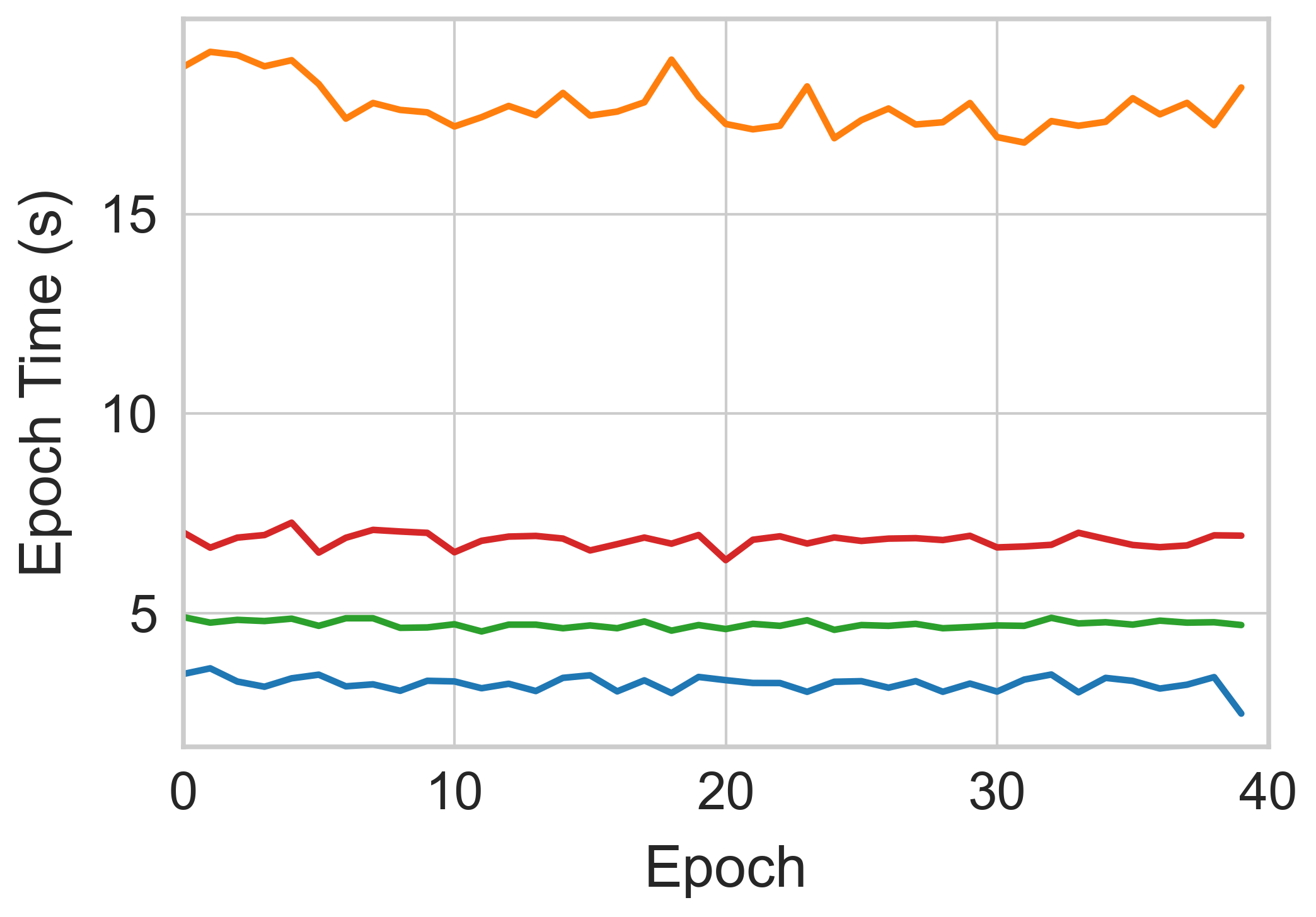}
    \caption{OGBN‑Products, batch size 3000}
    \label{fig:ogbn_3000}
  \end{subfigure}

  \vspace{1mm}

  \begin{subfigure}[b]{0.27\textwidth}
    \centering
    \includegraphics[width=\linewidth]{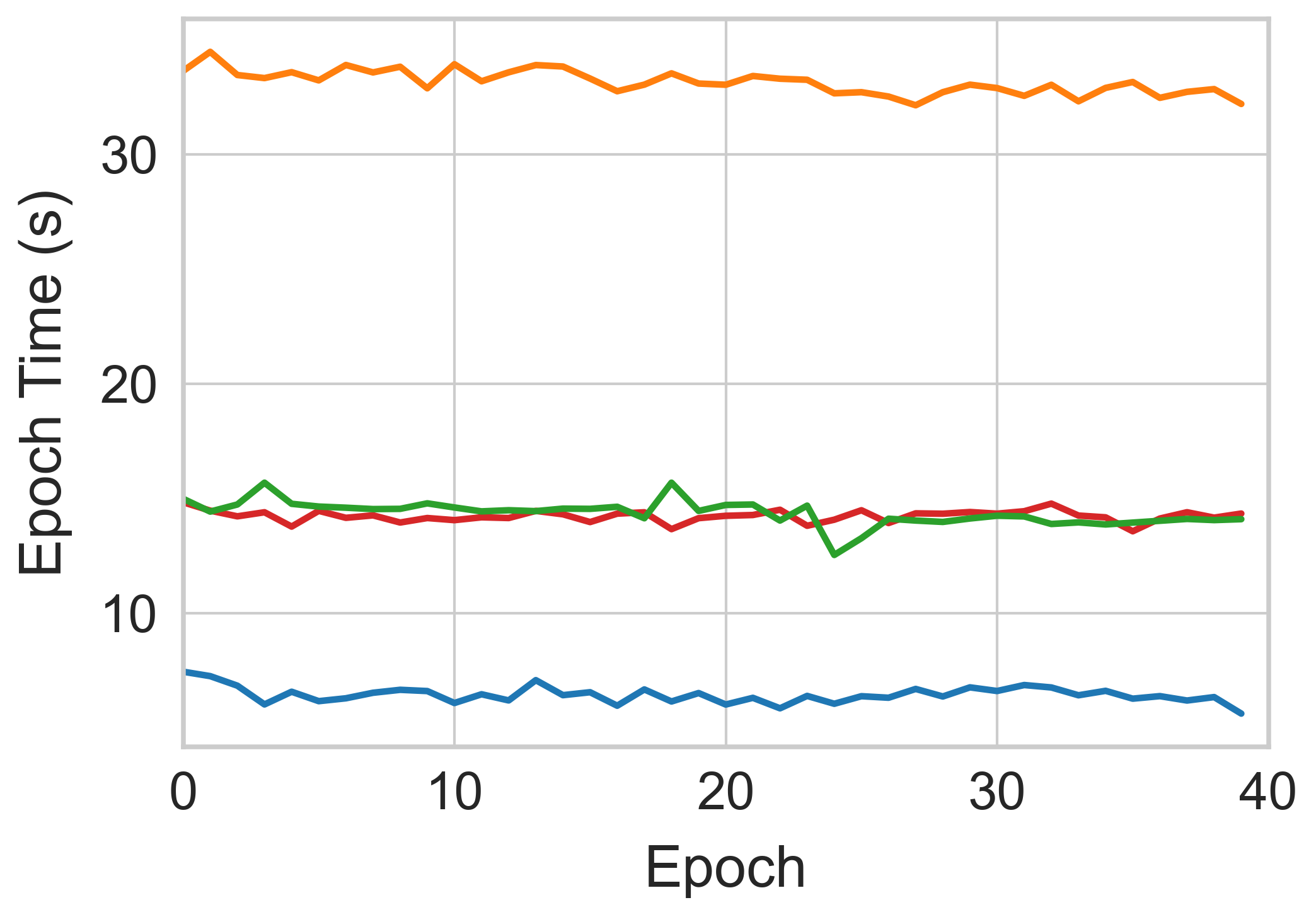}
    \caption{Reddit, batch size 1000}
    \label{fig:reddit_1000}
  \end{subfigure}\hfill
  \begin{subfigure}[b]{0.27\textwidth}
    \centering
    \includegraphics[width=\linewidth]{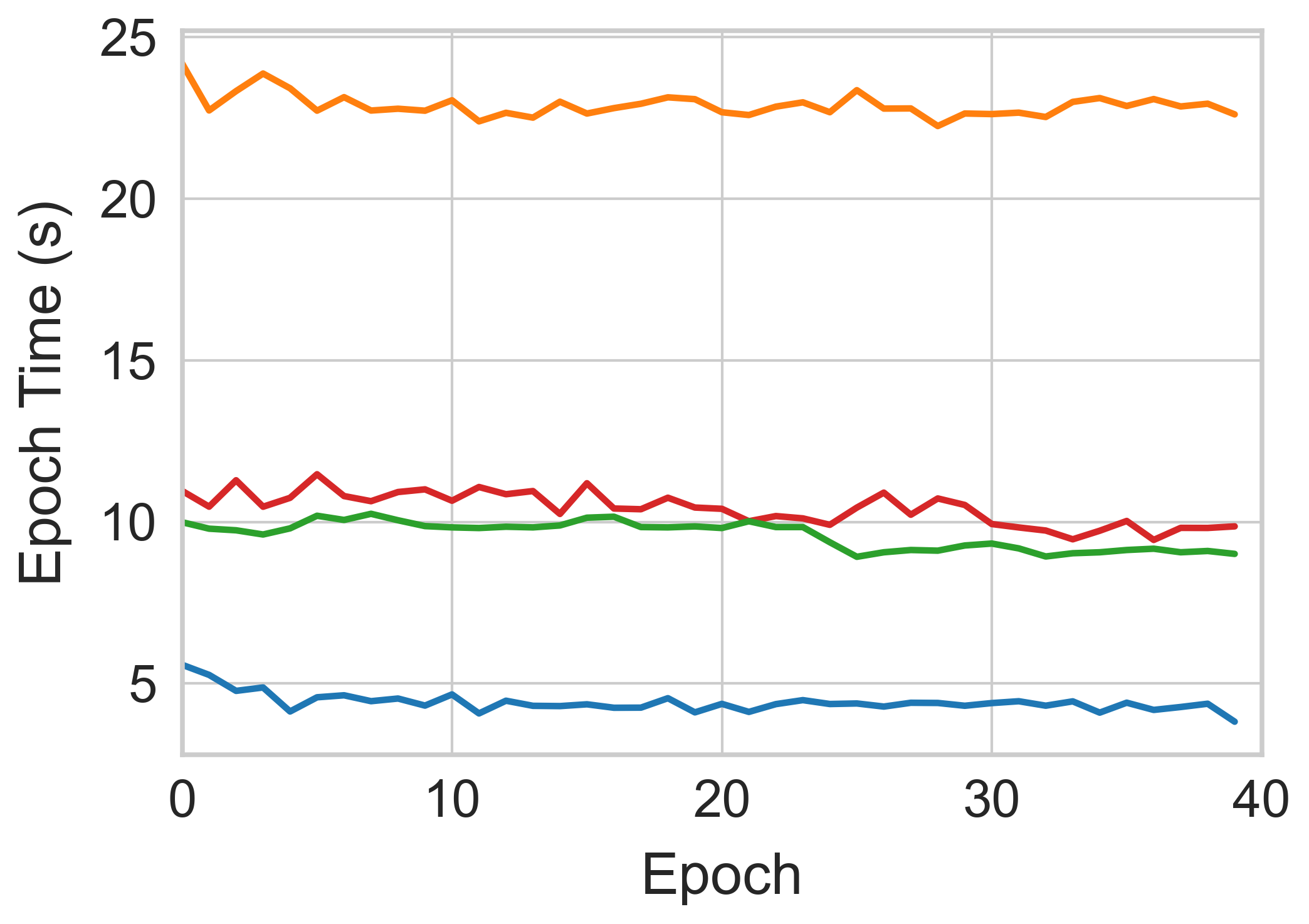}
    \caption{Reddit, batch size 2000}
    \label{fig:reddit_2000}
  \end{subfigure}\hfill
  \begin{subfigure}[b]{0.27\textwidth}
    \centering
    \includegraphics[width=\linewidth]{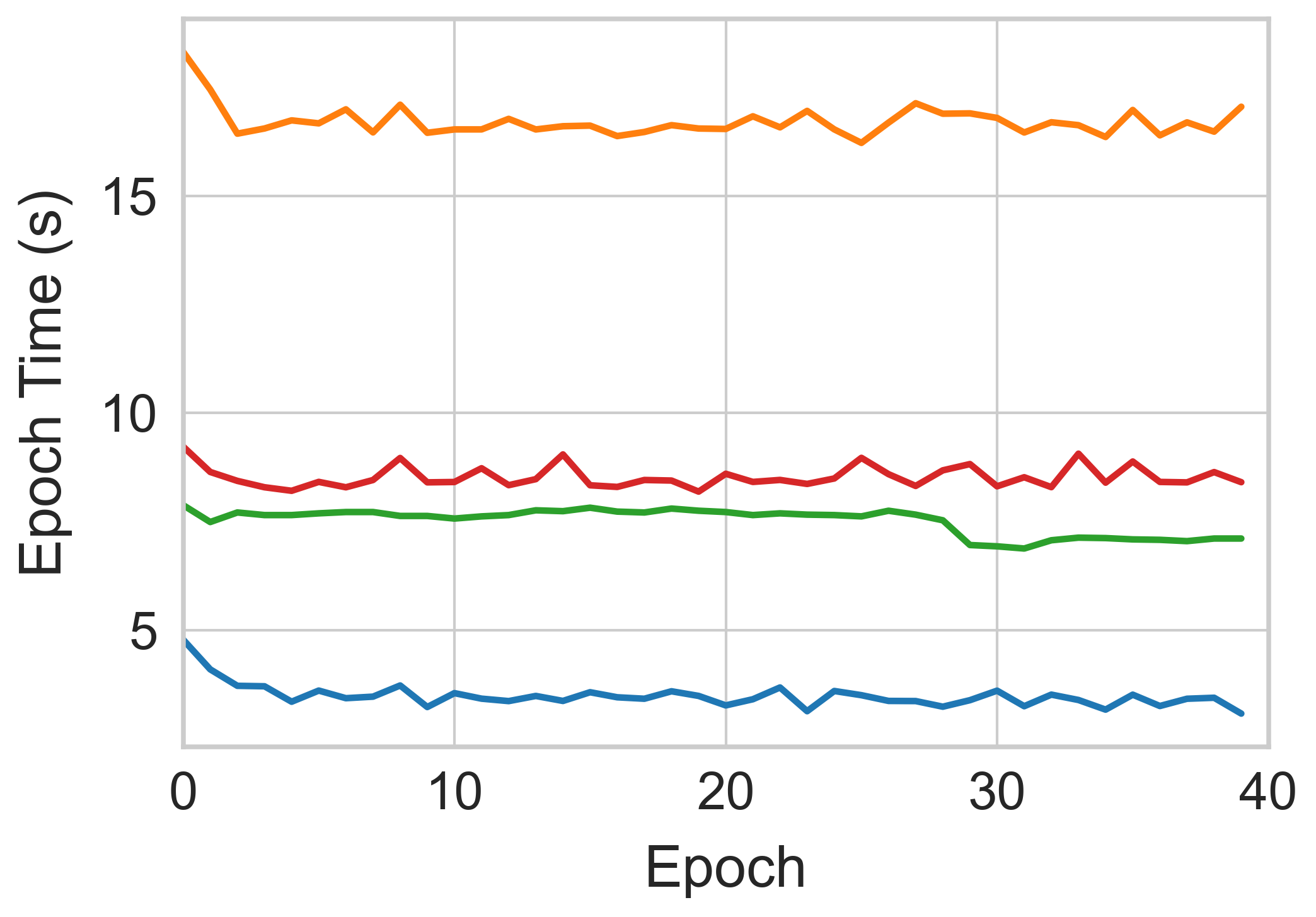}
    \caption{Reddit, batch size 3000}
    \label{fig:reddit_3000}
  \end{subfigure}

  \vspace{-1mm}

  \caption{Epoch‑time comparison across batch sizes on OGBN‑Products (top) and Reddit (bottom)}
  \label{fig:epoch_time_comparison}
\end{figure*}
\subsection{Experimental Setup}

We perform the experiments on the Reddit (232K nodes, 114.8M edges) and OGBN-Products (2.4M nodes, 123.7M edges) graph datasets, two standard benchmarks for GNN model performance. The properties and statistics of the datasets are detailed in Table~\ref{tab:dataset-properties}.

\begin{table}[!htbp]
  \centering
  \caption{Graph and Partitioning Properties of Reddit and OGBN-Products}
  \label{tab:dataset-properties}
  \begin{tabularx}{\columnwidth}{@{\extracolsep{\fill}} l c c @{}}
    \toprule
    \textbf{Property} & \textbf{Reddit} & \textbf{OGBN-Products} \\
    \midrule
    \multicolumn{3}{@{}l}{\textit{Graph Statistics}} \\
    \midrule
    \hspace{2mm}Number of Nodes       & 232,965       & 2,449,029     \\
    \hspace{2mm}Number of Edges       & 114,848,857   & 123,718,280   \\
    \hspace{2mm}Average Degree        & $\sim$492     & $\sim$101     \\
    \hspace{2mm}Number of Classes     & 50            & 47            \\
    \hspace{2mm}Feature Dimension     & 602           & 100           \\
    \midrule
    \multicolumn{3}{@{}l}{\textit{Partitioning Scheme}} \\
    \midrule
    \hspace{2mm}METIS                 & \checkmark    & \checkmark    \\
    \hspace{2mm}Random                & \checkmark    & \checkmark    \\
    \bottomrule
  \end{tabularx}
\end{table}

Both datasets are node-classification tasks (e.g., 50 classes for Reddit, 47 classes for OGBN-Products) with input feature dimensions of 602 and 100, respectively. The graphs are partitioned with a Random partition algorithm~\cite{hamilton2017inductive}  and METIS that aims to optimize communication with a balanced edge-cut objective. These partition schemes allow for each machine to work with a partition. We allow one halo hop so that each partition’s storage can have the immediate neighbor of its owned node as a ghost node. This is a common practice to reduce communication overhead for one-hop neighborhoods. However,  such approaches cannot account for hops and neighborhoods beyond that. The graph datasets used in this work are large enough to benefit from distributed training and have distinct structural properties, such as Reddit being more homogenous, while OGBN-Products have power-law degree distribution, providing a good test for our approach. Each node in these datasets has a high-dimensional feature vector (dense attributes), thus validating the costly feature fetching operation. 

We compare our method with three other models - DistDGL GCN~\cite{zheng2020distdgl}, GraphSAGE-Random~\cite{hamilton2017inductive}, and GraphSAGE METIS~\cite{hamilton2017inductive}. The GCN implementation requires the most expensive feature fetching operations as it does not use neighborhood sampling. GraphSAGE-Random does not use any optimization at the partition phase, while GraphSAGE-METIS optimizes communication overhead by balancing edges using METIS at the partition phase. 

We use Chameleon Cloud~\cite{keahey2020lessons} GPU nodes to conduct the experiments which are specified in Table \ref{tab:node-specs}.

\begin{table}[htbp]
  \centering
  \caption{Compute Node specifications for RapidGNN training}
  \label{tab:node-specs}
  \begin{tabularx}{\columnwidth}{@{}l X @{}}
    \toprule
    \textbf{Component}   & \textbf{Specification}                                      \\
    \midrule
    Platform             & Chameleon Cloud                    \\
    Processor            & 2× Intel Xeon E5-2670 v3 (12 cores each, 48 threads total)  \\
    Memory               & 128\,GiB RAM                                                \\
    GPU                  & 2× NVIDIA Tesla P100                                        \\
    GPU Memory           & 16\,GiB per GPU                                             \\
    Storage              & 400--1000\,GB local SSD                                     \\
    Network              & 10\,Gbps Ethernet                                           \\
    Operating System     & Ubuntu 22.04 LTS          \\
    \bottomrule
  \end{tabularx}
\end{table}

The \name and SOTA models run on identical hardware and software environments and use the exact sampling fan-out configuration and hyper-parameters. \name is implemented as described in Section IV, with the cache size tuned from $n_{\text{hot}}=25$K nodes to $n_{\text{hot}}=200$K, corresponding to roughly the top 15\% of remote nodes in each case—determined via a short profiling run). For the prefetcher, we set queue length $Q=3$ batches to balance between latency hiding and memory footprint, which is subject to hardware capabilities and can be tuned according to the machine's configuration.

We train for multiple epochs in all experiments and report per-epoch performance metrics. We use the Nvidia NVML~\cite{nvml_lib} and psutils~\cite{psutil_lib} libraries to measure the CPU and GPU metrics during training. 

\subsection{Training Time and Throughput}

\paragraph{Epoch Time Speedup}

\name delivers substantial acceleration across the datasets and all batch sizes. Table~\ref{tab:speedup} reports the speedup factors relative to GCN, GraphSAGE-METIS, and GraphSAGE-Random. Averaged over six configurations, \name is \textbf{1.84$\times$} faster than GCN and \textbf{2.10$\times$} faster than GraphSAGE-METIS. The most significant single gain—\textbf{5.76$\times$} over GraphSAGE-Random—occurs on OGBN-Products at batch size 1000, while Reddits sees up to \textbf{5.18$\times$} over GraphSAGE-Random at batch size 2000. Figure~\ref{fig:epoch_time_comparison} shows the detailed per-epoch time for these configurations. It shows \name consistently outperforms GCN, GraphSAGE-Random, and GraphSAGE-METIS throughout the whole training. 
The improvement comes from dramatically reducing the waiting time for on-demand feature fetching and using the prefetcher to feed the features to training. GraphSAGE-Random performs the worst as without any heuristic to guide the partitioning, almost every single edge can be a cross-partition edge, thus incurring massive communication. The initial spike consistently seen across all instances of \name is due to the initial warm-up phase when the prefetcher is empty and the dip at the tail is due to the absence of any more batches to prefetch.

\begin{table}[!htbp]
  \centering
  \footnotesize
  \caption{Speedup of \name over SOTA models}
  \label{tab:speedup}
  \begin{tabular}{llccc}
    \toprule
    \textbf{Dataset}     & \textbf{Batch Size} & \textbf{GCN} & \multicolumn{2}{c}{\textbf{GraphSAGE}} \\
    \cmidrule(lr){4-5}
    & & & \textbf{METIS} & \textbf{Random} \\
    \midrule
    OGBN-Products        & 1000 & 1.50 & 1.74 & 5.76 \\
                         & 2000 & 1.56 & 1.79 & 5.72 \\
                         & 3000 & 1.46 & 2.11 & 5.47 \\
    Reddit               & 1000 & 2.23 & 2.21 & 5.14 \\
                         & 2000 & 2.17 & 2.36 & 5.18 \\
                         & 3000 & 2.16 & 2.45 & 4.80 \\
    \midrule
    \multicolumn{2}{l}{\textbf{Average Speedup}} 
                         & \textbf{1.84×} & \textbf{2.10×} & \textbf{5.34×} \\
    \bottomrule
  \end{tabular}
\end{table}

\begin{figure}[!htbp]
  \centering
  \begin{subfigure}[b]{0.45\textwidth}
    \includegraphics[width=\linewidth]{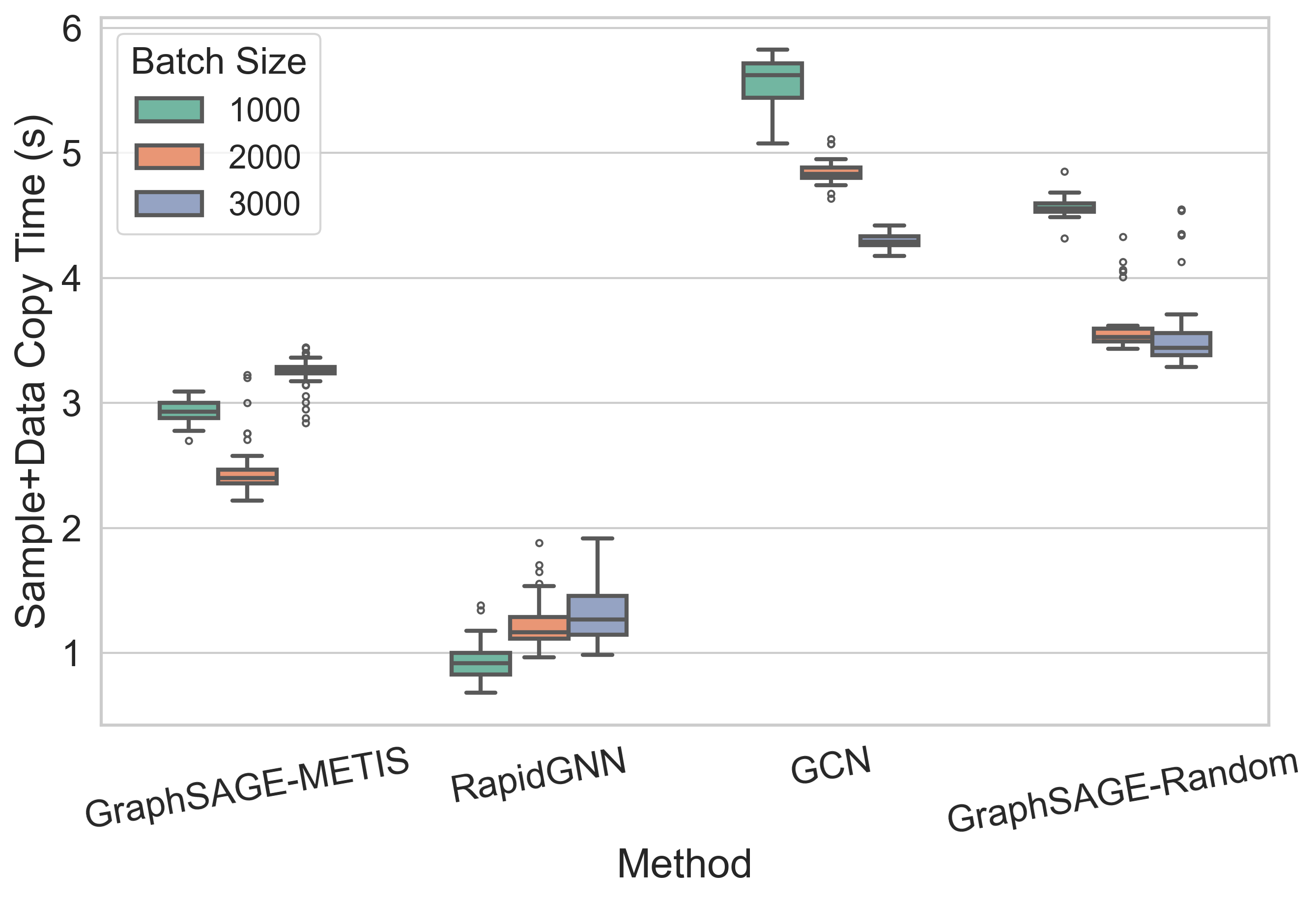}
    \caption{OGBN-Products}
    \label{fig:sample_data_ogbn}
  \end{subfigure}\hfill
  \begin{subfigure}[b]{0.45\textwidth}
    \includegraphics[width=\linewidth]{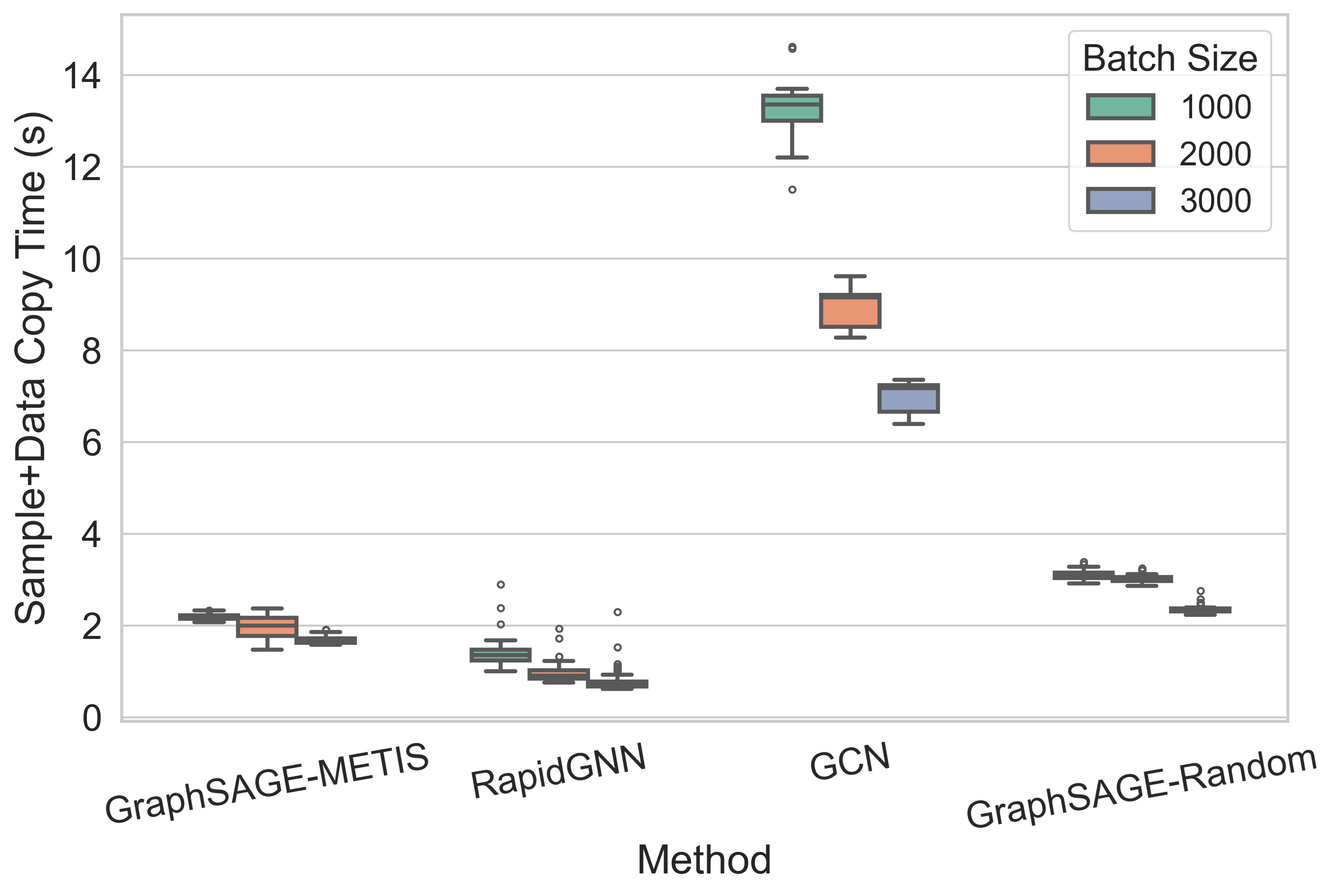}
    \caption{Reddit}
    \label{fig:sample_data_reddit}
  \end{subfigure}
  \caption{Sampling + data copy time distributions for GCN, \name, and GraphSAGE variants.}
  \label{fig:sample_data}
  \vspace{-1mm}
\end{figure}

\paragraph{Sampling + Data Copy Time}
To understand the results found in the previous observation, we measure the time spent to sample and copy the data required for training to the device. Figure~\ref{fig:sample_data} presents boxplots of the sampling + data copy phase for each method on OGBN-Products (panel a) and Reddit (panel b). \name consistently achieves the lowest median and tightest interquartile range: under 1s on OGBN-Products and below 1.4s on Reddit across all batch sizes. By contrast, GCN’s median exceeds 5s (OGBN-Products) and climbs from 8s to over 13s (Reddit), while GraphSAGE‐based methods remain in the 2–3s range with broader variance and frequent high‐latency outliers.

On Reddit at batch size 1000, \name reduces mean copy time by 89.2\% versus GCN and by 34.6\% versus GraphSAGE-METIS; at batch size 3000 the reductions are 88.3 \% and 51.3\%, respectively. On OGBN-Products (batch size 1000), feature copy time drops by 83.2\% versus GCN and 68\% versus GraphSAGE-METIS. Averaged over all six cases, \name cuts sampling + data copy overhead by 82.3\% against GCN and 52.2\% against GraphSAGE-METIS, directly contributing to the epoch‐time gains above. The tight distribution observed in our implementation indicates that it has removed much of the randomness and spikes from the data loading phase by steadily supplying the features to the training process. Nearly every batch is ready when required, leading to consistently low latency.

\subsection{Communication Reduction and Feature Reuse}

\begin{figure}[t]
    \centering
    \includegraphics[width=0.44\textwidth]{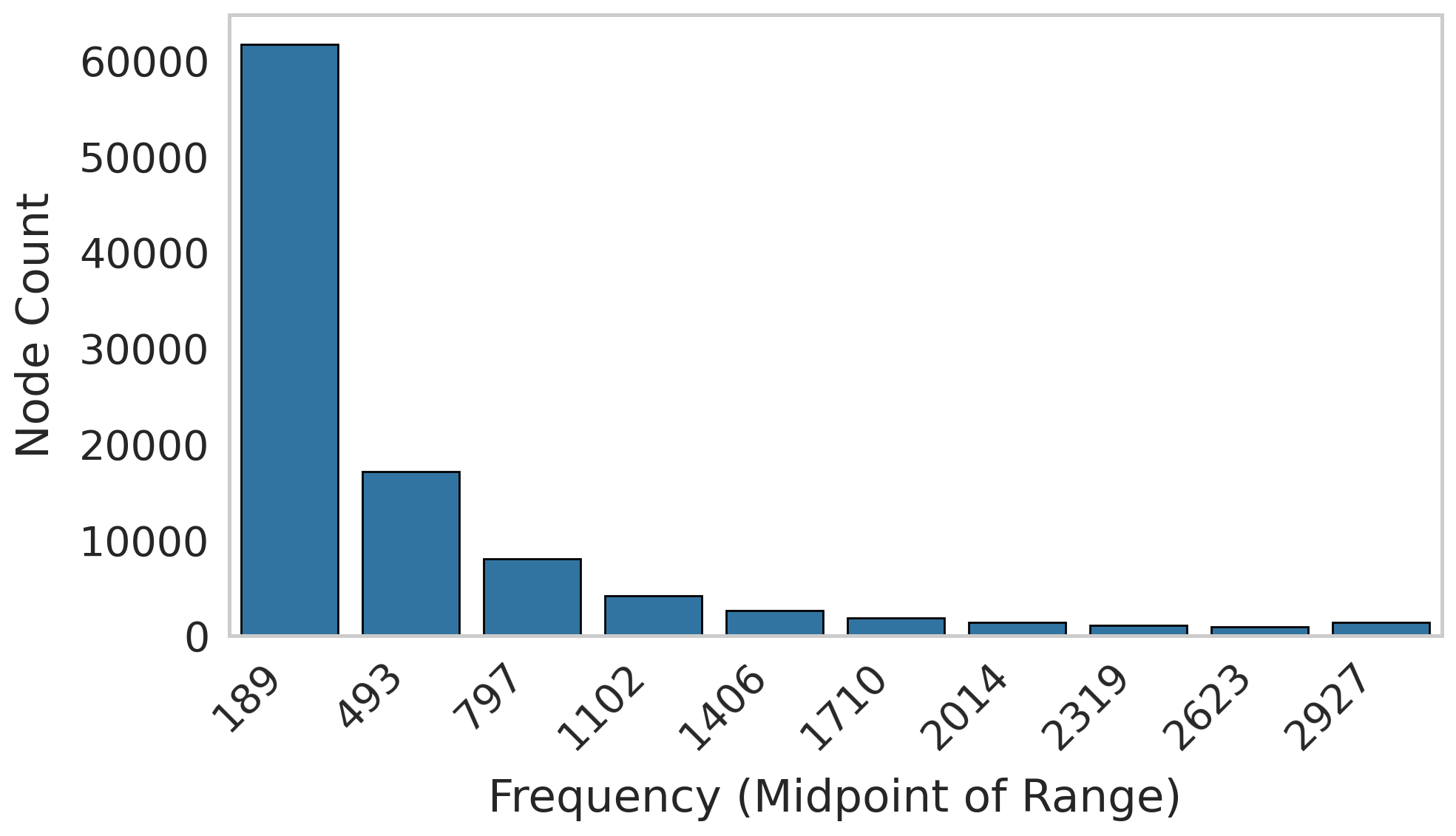}
    \caption{Frequency distribution of remote feature accesses per node (midpoint of range on the x-axis). Most nodes are fetched only a handful of times, indicating a long-tail reuse pattern.}
    \label{fig:frequency}
    \vspace{-1mm}
\end{figure}

Then, we analyze \name's communication efficiency in reducing the number of calls for remote feature fetching and the volume of data transferred over the network. We tally the frequency of access of remote nodes' features and cache the most frequently used remote nodes in an epoch according to the frequency distribution. Figure~\ref{fig:frequency} shows the distribution of how often each remote node’s feature is fetched during training (binned by midpoint frequency). Roughly 60,000 nodes are accessed with very low frequency (midpoint = 189), and fewer than 5,000 nodes exceed a frequency of 1400, demonstrating that a small fraction of “hot” nodes account for most remote requests.

\begin{figure}[!htbp]
    \centering
    \includegraphics[width=0.43\textwidth]{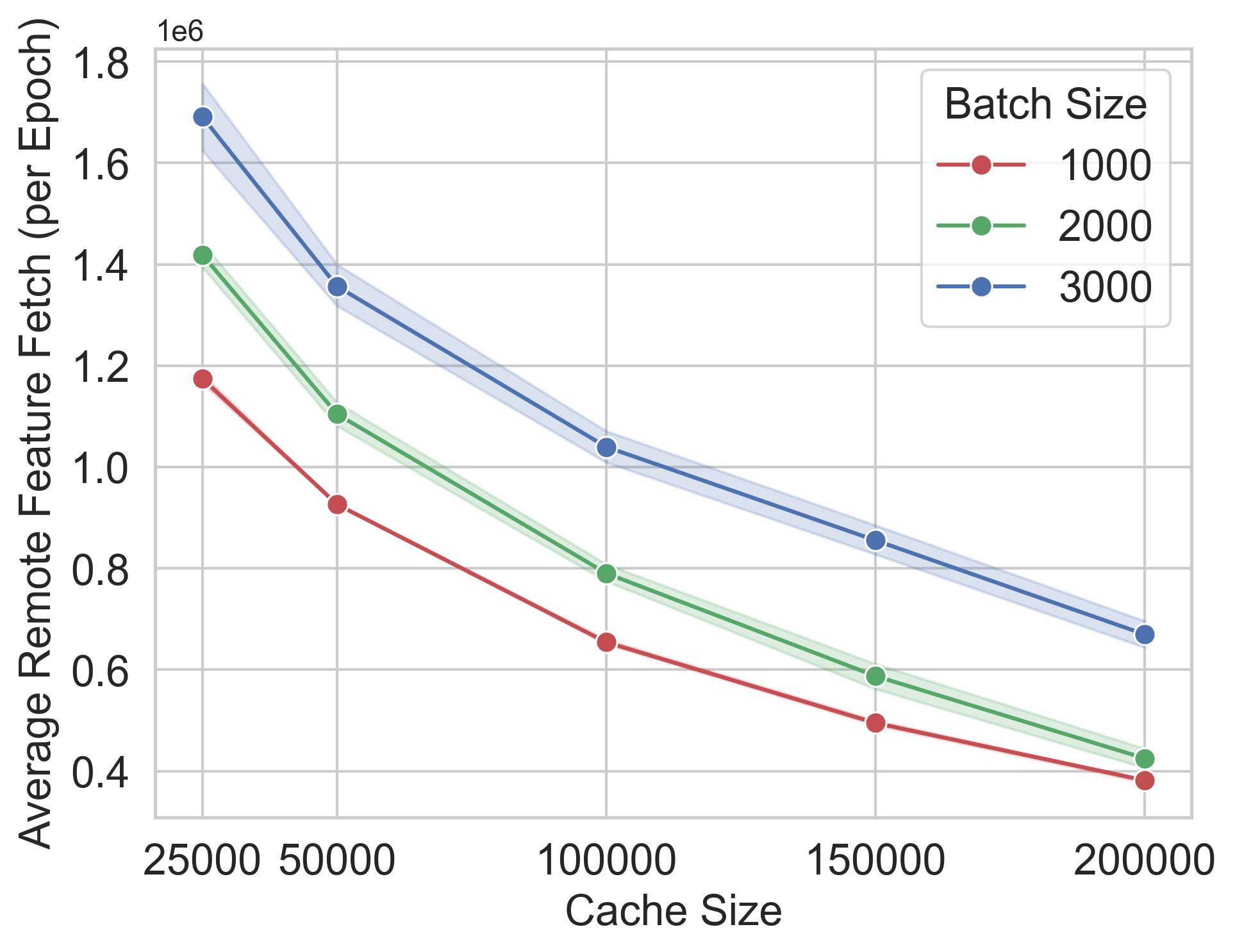}
    \caption{Average number of remote feature fetches per epoch versus cache size}
    \label{fig:avg_rpc}
    \vspace{-1mm}
\end{figure}

Figure~\ref{fig:avg_rpc} quantifies the impact of cache size on remote‐fetch volume. At a small cache of 25,000 nodes, large batches (3000) incur over 1.7 million remote fetches per epoch, whereas smaller batches (1000) still see roughly 1.2 million. As cache size grows to 200,000, fetch counts drop to ~0.35 million (batch 1000), ~0.45 million (batch 2000), and ~0.80 million (batch 3000). This nearly linear decrease confirms that caching the top‐frequency nodes—identified in Figure~\ref{fig:frequency}—substantially reduces communication overhead, with larger caches yielding diminishing returns as the long tail thins out.


\begin{figure}[t]
    \centering
    \begin{subfigure}{0.24\textwidth}
        \includegraphics[width=\linewidth]{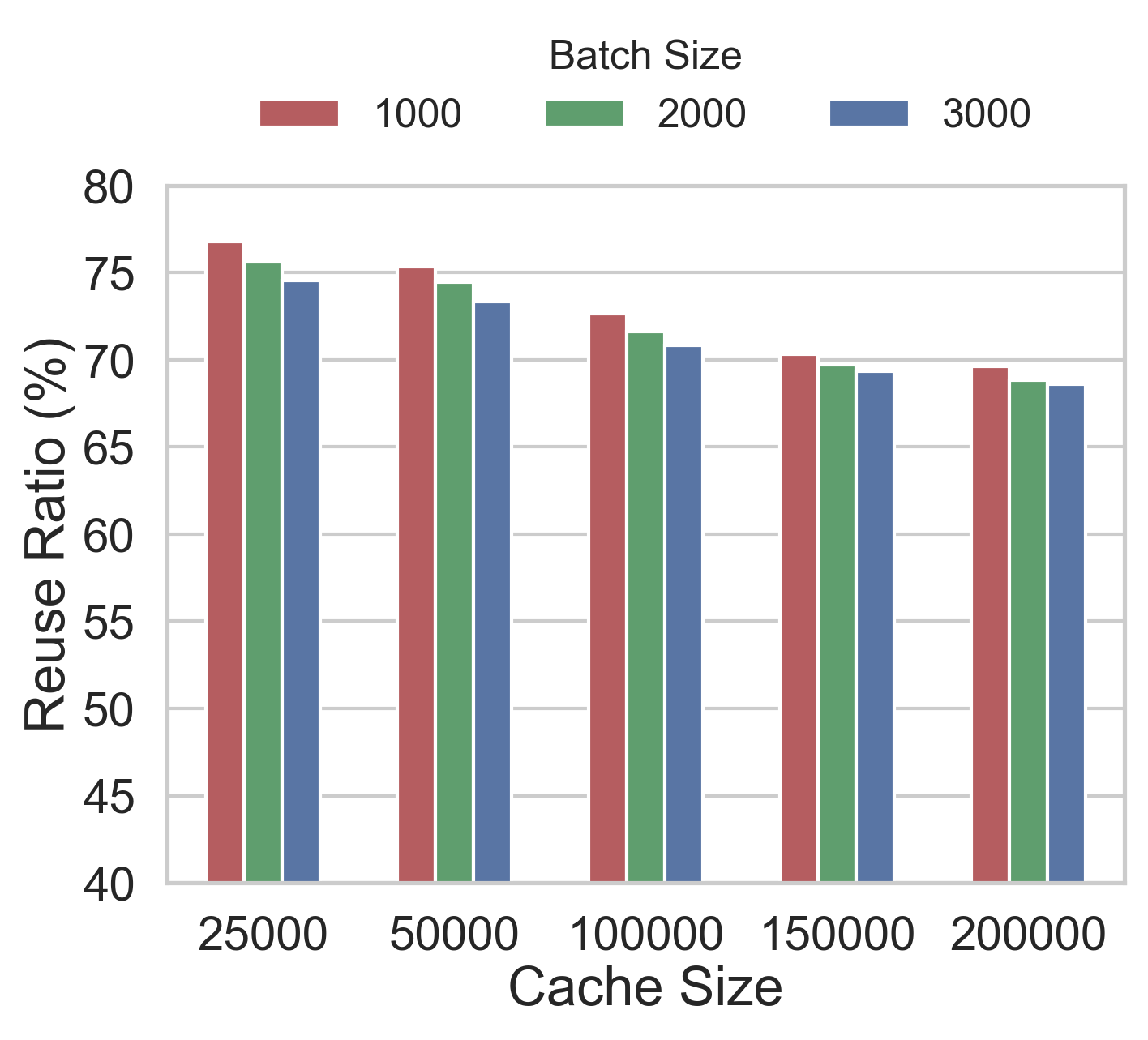}
        \caption{Reuse Ratio across Cache Sizes}
        \label{fig:reuse_ratio}
    \end{subfigure}\hfill
    \begin{subfigure}{0.24\textwidth}
        \includegraphics[width=\linewidth]{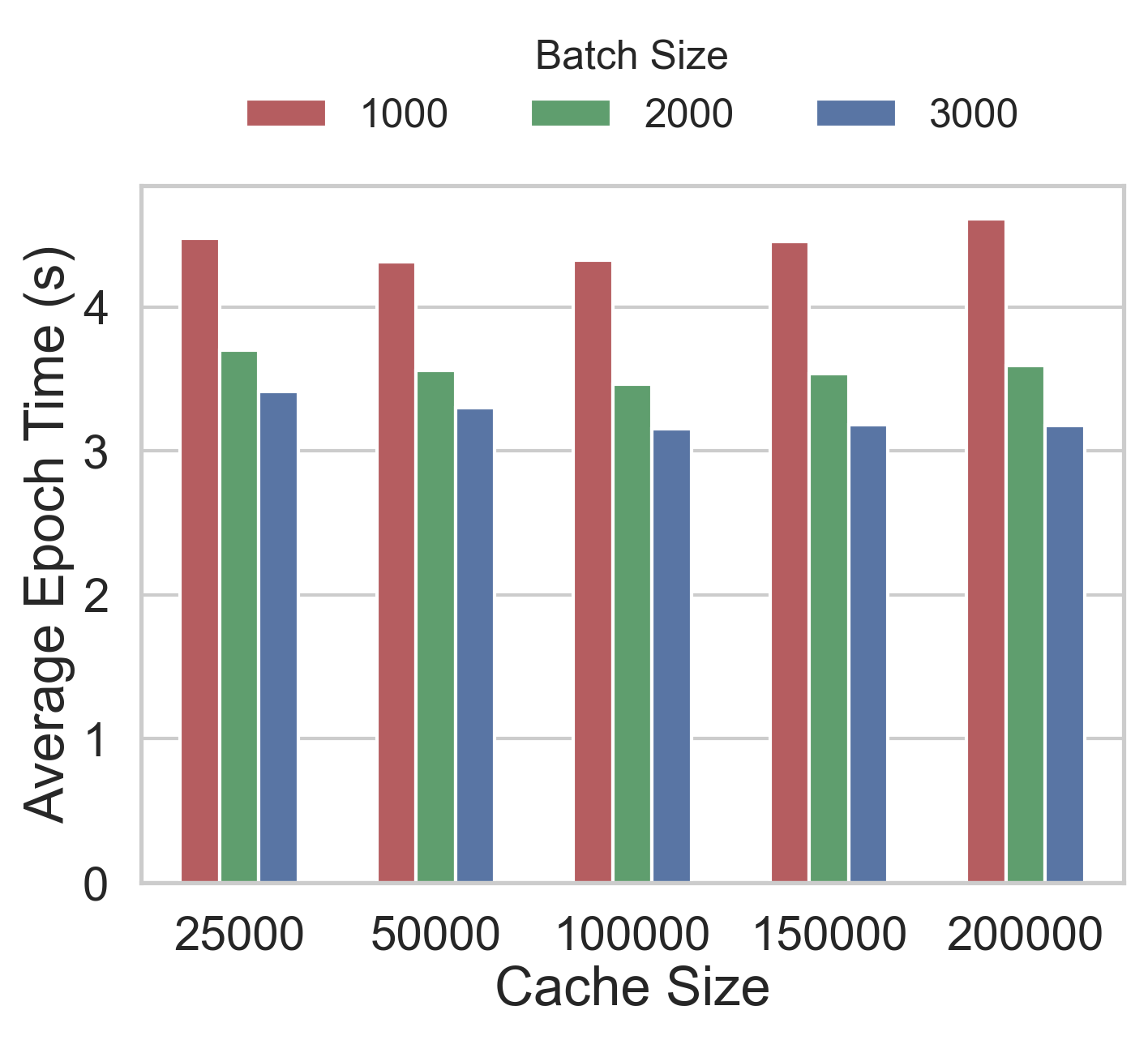}
        \caption{Average Epoch Time across Cache Sizes}
        \label{fig:avg_epoch_time}
    \end{subfigure}
    \caption{Comparison of Reuse Ratio and Average Epoch Time for Different Cache Sizes}
    \label{fig:reuse_vs_epoch}
    \vspace{-2mm}
\end{figure}

Figure~\ref{fig:reuse_vs_epoch}(a) confirms that a very small hot‐node cache captures the lion’s share of reuse: with only 25,000 entries (\(\approx 1\%\) of the graph), the reuse ratio exceeds 78\% for batch size 1000 and 73\% for batch size 3000. Increasing the cache to 50,000 or 100,000 nodes yields only marginal changes in reuse (\(\pm 2\%\)), and even a 200,000-node cache—eight times larger—only shifts the reuse-ratio by another 7–8\%. The decrease in reuse ratio reflects the long-tail distribution of feature requests: after caching the core “hot” set, each extra node contributes very little additional reuse.

Despite these shifts in reuse, average epoch time (Fig.~\ref{fig:avg_epoch_time}) remains essentially flat across all cache sizes and batch configurations, varying by less than 5\% in most cases. This stability indicates that further cache expansion does not translate into measurable runtime gains once the core hot-node set is stored. In practice, one can provision a cache of 50,000–100,000 nodes to achieve near-peak reuse while minimizing memory overhead without losing epoch throughput. However, that stability is mainly due to optimization at the prefetcher level, which hides the latency of fetching the features behind the training time. Therefore, the primary contribution of caching is in reducing the number of redundant feature fetching instead of directly reducing the epoch time, as shown in Figure \ref{fig:avg_rpc}.

\begin{figure}[!hbtp]
  \centering
  \includegraphics[width=0.5\columnwidth]{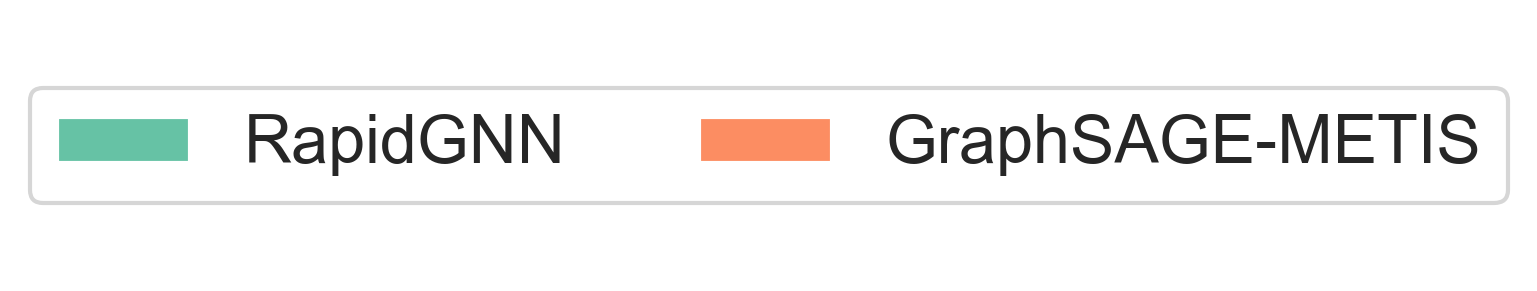}\\[1ex]
  \begin{subfigure}[b]{0.5\columnwidth}
    \centering
    \includegraphics[width=4cm,height=5cm]{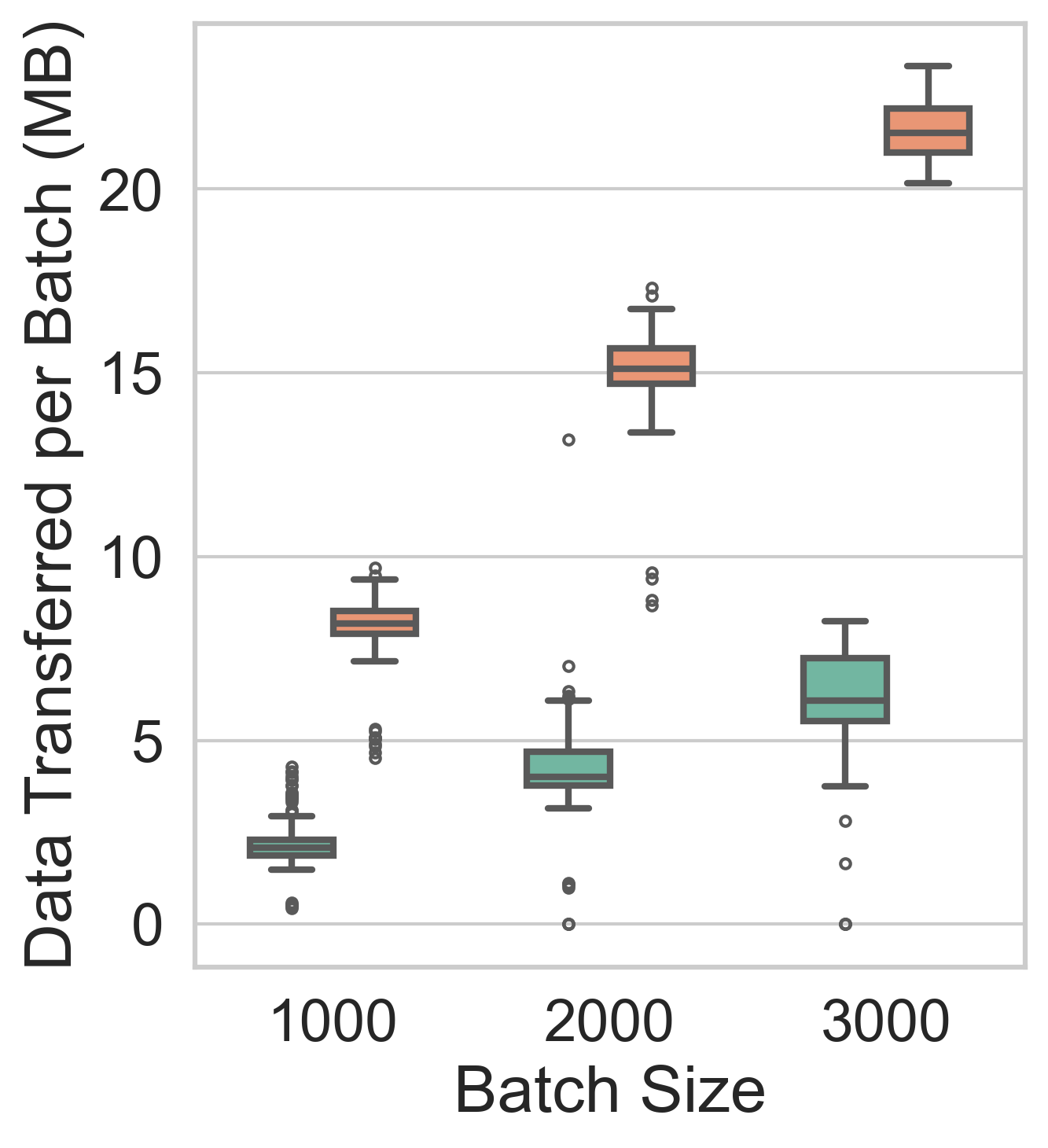}
    \caption{Data transferred}
    \label{fig:sub1}
  \end{subfigure}\hfill
  \begin{subfigure}[b]{0.5\columnwidth}
    \centering
    \includegraphics[width=4cm,height=5cm]{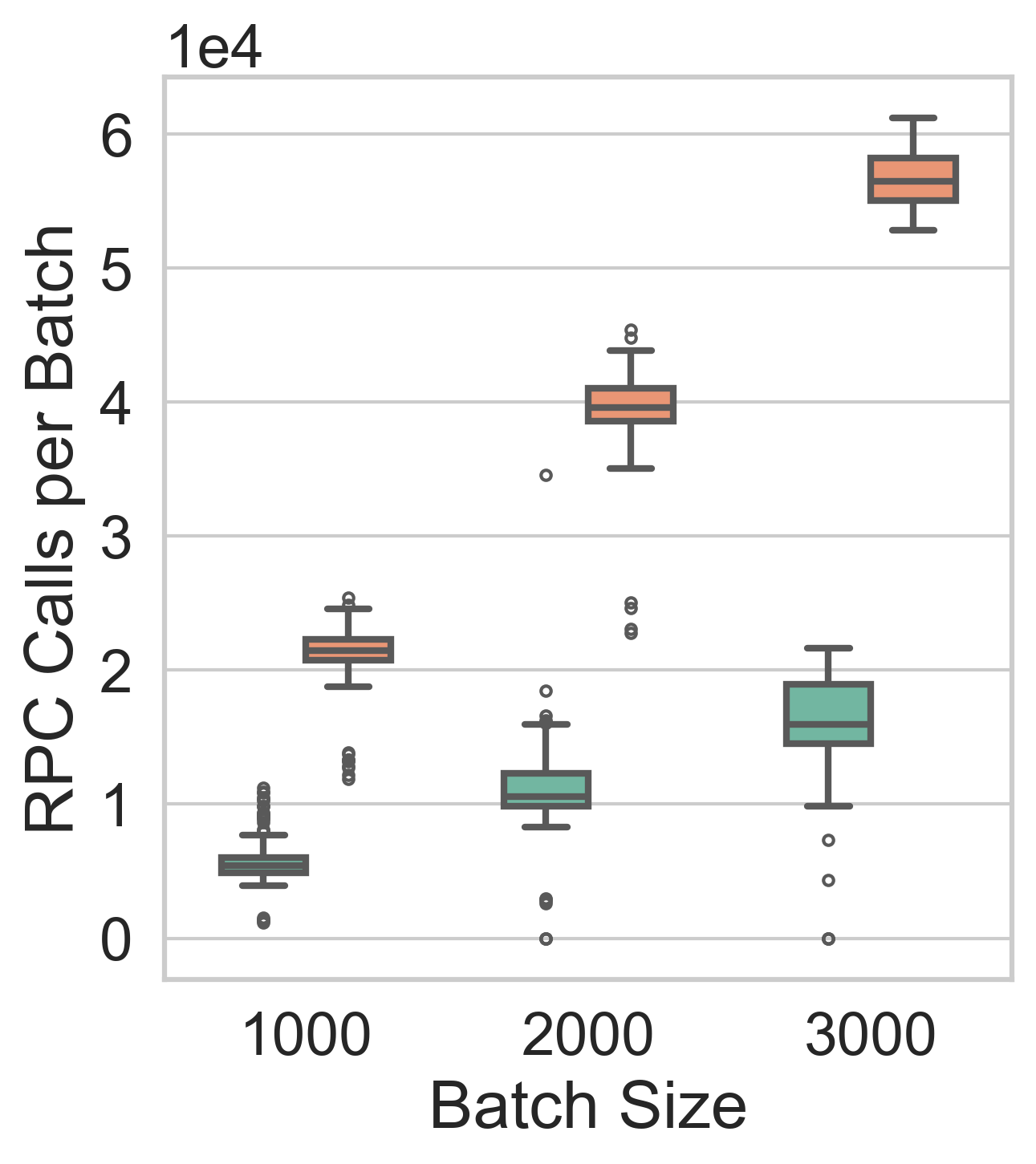}
    \caption{RPC calls}
    \label{fig:sub2}
  \end{subfigure}

  \caption{Comparison of per‐batch data and RPC calls.}
  \label{fig:side_by_side}
  \vspace{-3mm}
\end{figure}

We also instrument the system to count the number of RPC remote feature fetch calls and data transferred per batch. We mainly compare this against the GraphSAGE-METIS as it is the most superior out of the SOTA models from the previous evaluation of epoch time and throughput. \name demonstrates a significant reduction in the average number of RPC feature calls per batch and the amount of data transferred, which is averaged over multiple batch sizes over 40 epochs. Figure \ref{fig:sub1} and Figure \ref{fig:sub2} compare the Data transferred per batch and the Number of RPC feature calls per batch, respectively, for \name and GraphSAGE-METIS. 

\name reduces the volume of transferred data and the number of RPC calls by 4$\times$. We can also see that the number of RPC feature calls for remote nodes directly correlates to the volume of data transferred from these two figures. Essentially, by reusing the remote nodes' features and using the precomputed access pattern, the cache can reduce the number of dispatched feature calls to the remote feature store by 4$\times$, thus reducing the data volume.

\subsection{Resource Usage and Energy Efficiency}
Along with reducing communication overhead and training time, \name also improves energy efficiency. We measure the energy consumption for batch size 1000 for OGBN-Products dataset over 40 epochs and averaged it in Table~\ref{tab:graphcap_vs_graphsage}.
\vspace{-3mm}
\FloatBarrier

\begin{table}[!htbp]
  \centering
  \scriptsize
  \caption{Performance comparison between \name and GraphSAGE-METIS.}
  \label{tab:graphcap_vs_graphsage}
  \resizebox{\columnwidth}{!}{%
  \begin{tabular}{lccc}
    \toprule
    \textbf{Metric} & \textbf{\name} & \textbf{GraphSAGE-METIS} & \textbf{Difference} \\
    \midrule
    RPC Calls          & 522{,}230         & 2{,}129{,}287      & $\sim$4$\times$ fewer \\
    Data Transferred   & 199\,MB           & 812\,MB            & $\sim$4$\times$ less  \\
    CPU Memory        & 5.15\,GB          & 2.68\,GB           & $\sim$2$\times$ higher\\
    GPU Energy         & 376\,J            & 487\,J             & 23\% less         \\
    Total Energy      & 385\,J            & 491\,J             & 22\% less         \\
    \bottomrule
  \end{tabular}%
  }
\end{table}

Though the caching of the features increases the CPU memory usage, \name consumes about 376J of GPU energy per epoch compared to 487J in baseline (\textbf{23\%} reduction). The total system energy (including CPU) showed a similar \textbf{22\%} improvement. This stems from two factors: Shorter execution time – the faster the training completes an epoch, the less time the hardware draws power; and Less active communication – network interfaces and CPU cores spend less time busy-waiting or handling RPCs, which lowers their energy usage. By cutting redundant work, \name speeds up training and translates those savings into lower energy consumption.

\subsection{Convergence Evidence}
\begin{figure}[!t]
  \centering

  \includegraphics[width=0.45\textwidth]{figures/graphs/epochtime/separate_legend.png}

  \begin{subfigure}[b]{0.48\columnwidth}
    \includegraphics[width=\linewidth]{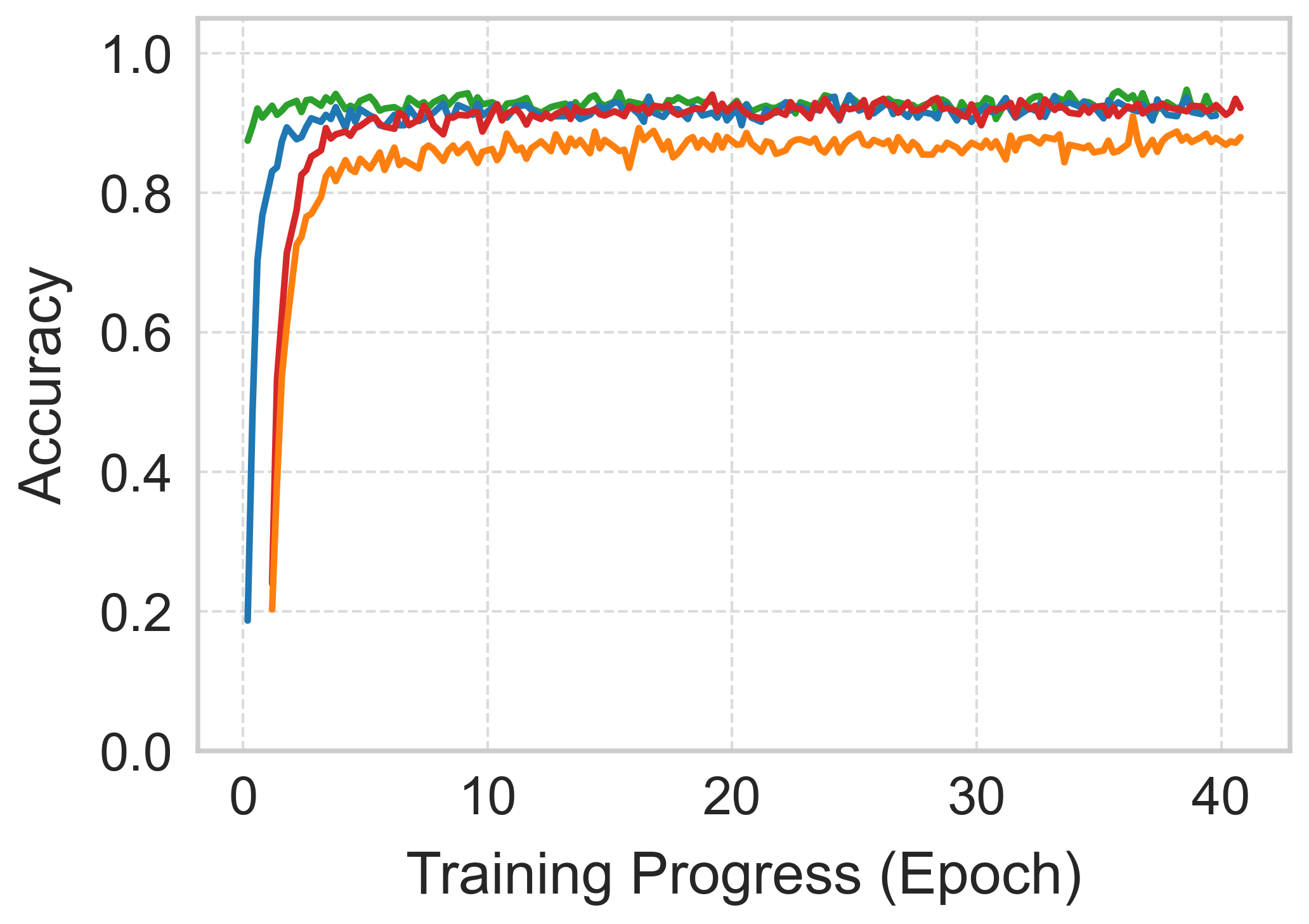}
    \caption{OGBN‑Products, batch size 1000}
    \label{fig:acc_ogbn_1000}
  \end{subfigure}\hfill
  \begin{subfigure}[b]{0.48\columnwidth}
    \includegraphics[width=\linewidth]{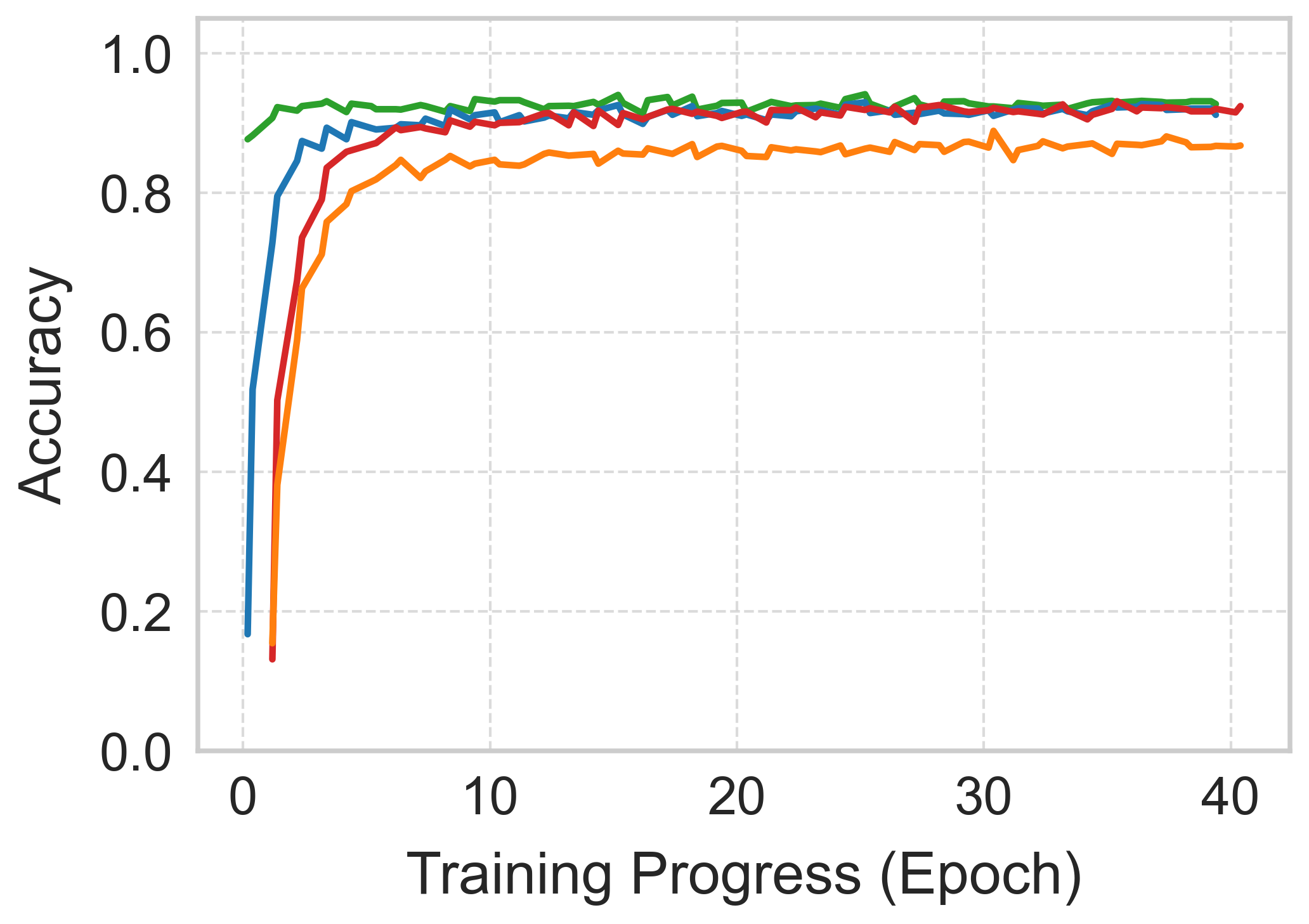}
    \caption{OGBN‑Products, batch size 2000}
    \label{fig:acc_ogbn_2000}
  \end{subfigure}

  \vspace{1mm}

  \begin{subfigure}[b]{0.48\columnwidth}
    \includegraphics[width=\linewidth]{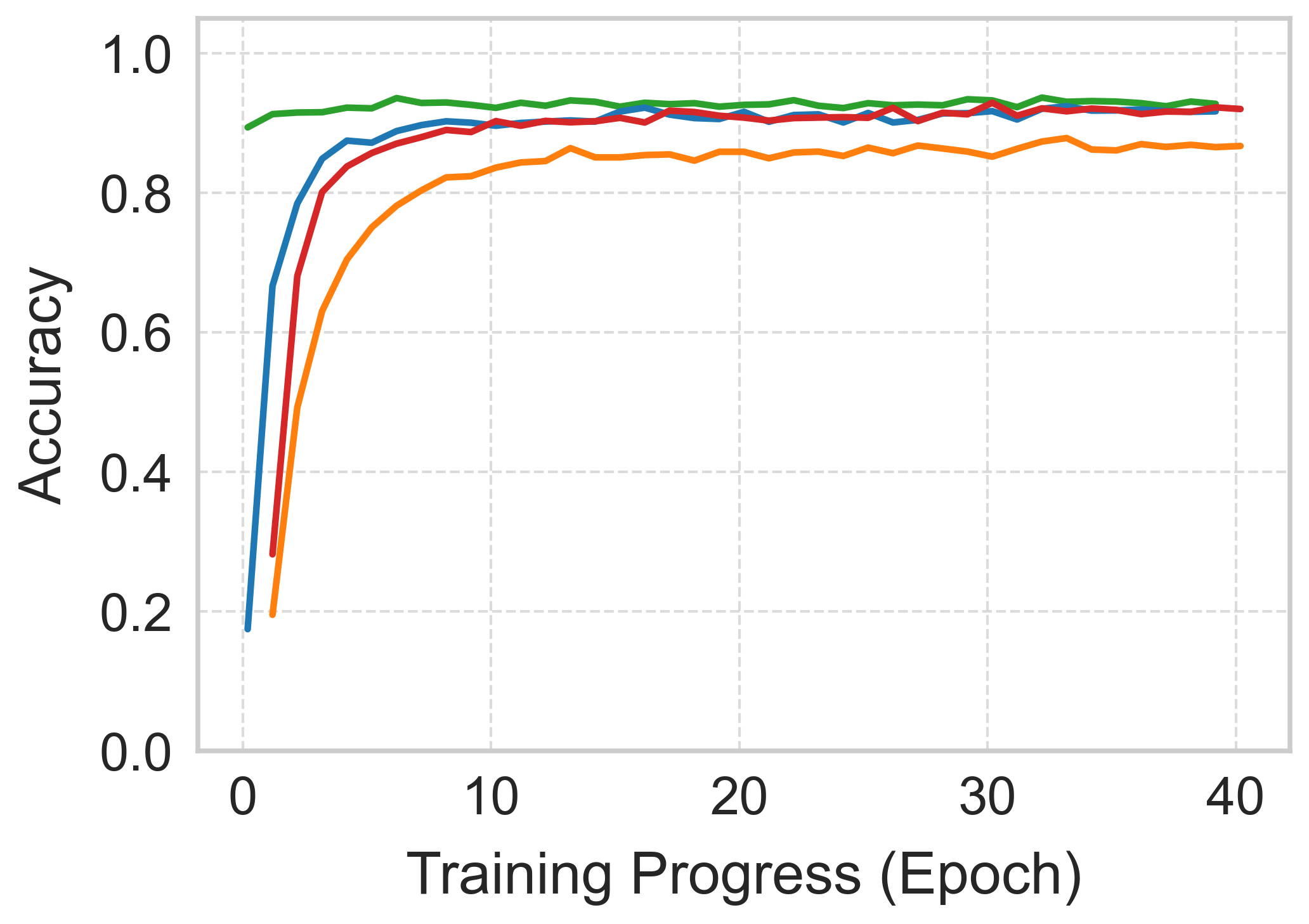}
    \caption{OGBN‑Products, batch size 3000}
    \label{fig:acc_ogbn_3000}
  \end{subfigure}\hfill
  \begin{subfigure}[b]{0.48\columnwidth}
    \includegraphics[width=\linewidth]{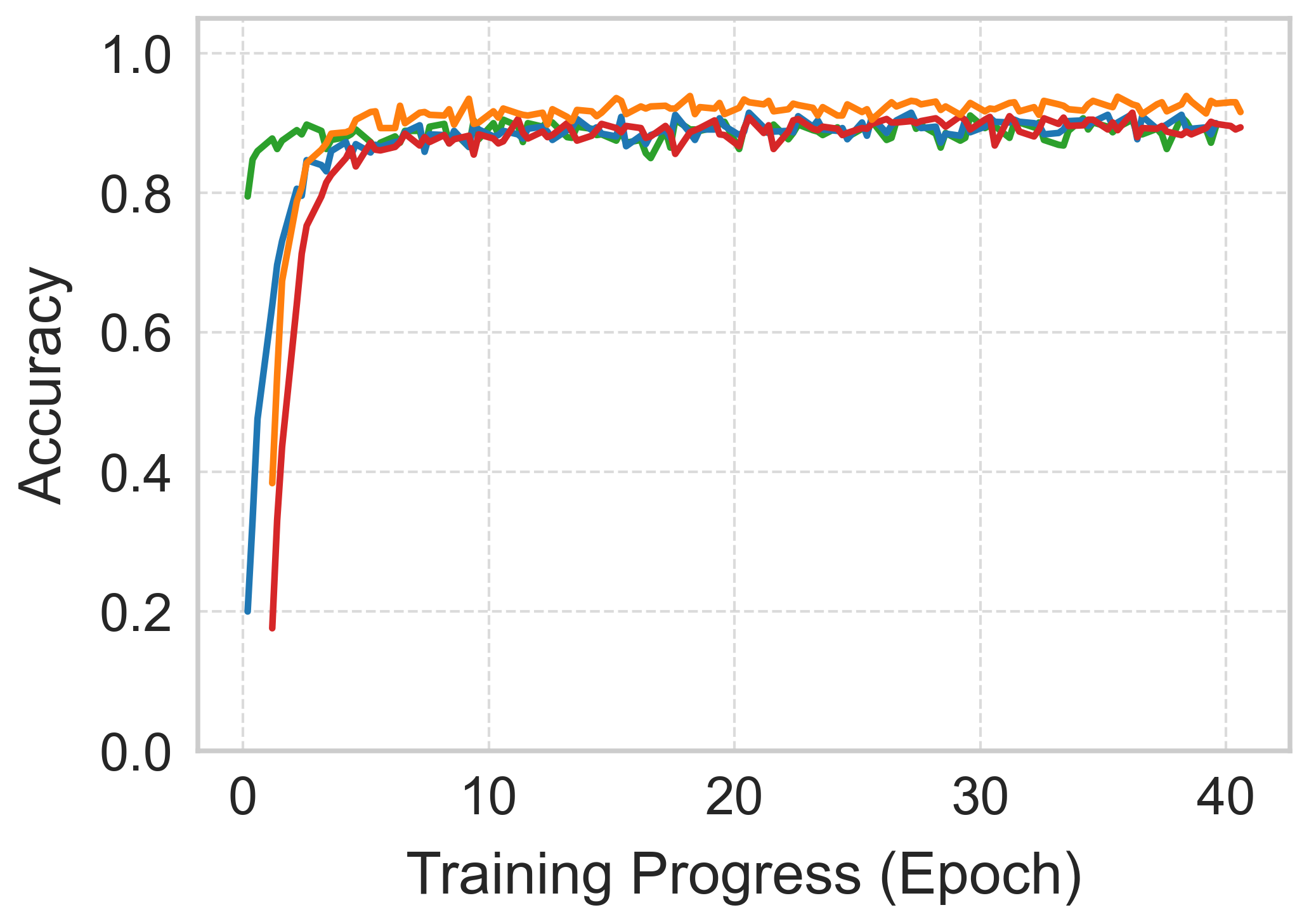}
    \caption{Reddit, batch size 1000}
    \label{fig:acc_reddit_1000}
  \end{subfigure}

  \vspace{1mm}

  \begin{subfigure}[b]{0.48\columnwidth}
    \includegraphics[width=\linewidth]{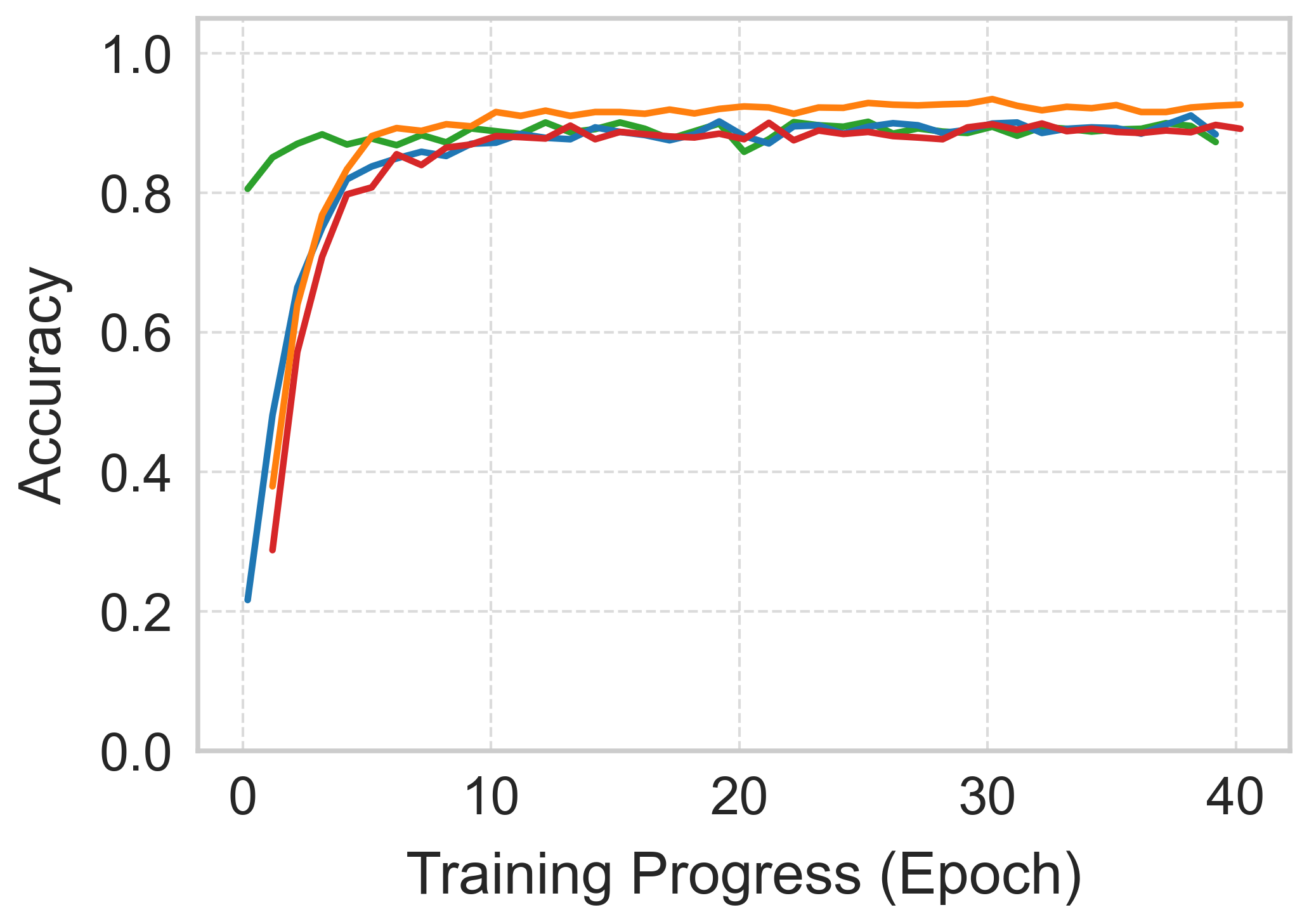}
    \caption{Reddit, batch size 2000}
    \label{fig:acc_reddit_2000}
  \end{subfigure}\hfill
  \begin{subfigure}[b]{0.48\columnwidth}
    \includegraphics[width=\linewidth]{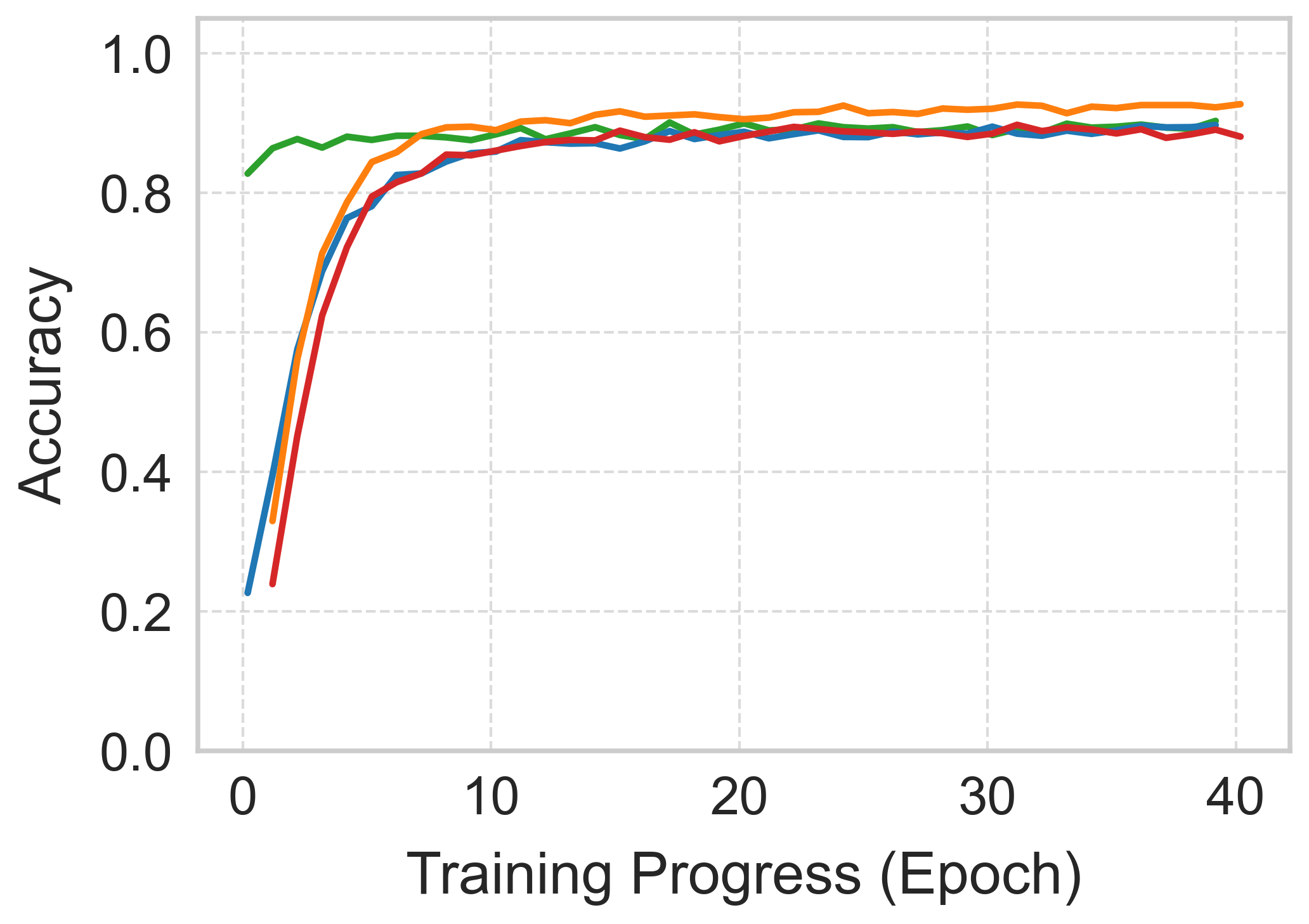}
    \caption{Reddit, batch size 3000}
    \label{fig:acc_reddit_3000}
  \end{subfigure}

  \vspace{-1.5mm} 
  \caption{Training accuracy across batch sizes on OGBN‑Products (top three) and Reddit (bottom three).}
  \label{fig:accuracy_comparison}
  \vspace{-4mm}
\end{figure}

To verify that our fixed‐seed sampling, caching, and asynchronous prefetching preserve standard SGD convergence, we compare epoch‐wise training accuracy of \name against the baselines in Figure~\ref{fig:accuracy_comparison}.

In all six configurations, \names’s accuracy curves rapidly rise and plateau at the same level as the baselines. We observe no signs of slowed convergence or increased variance due to deterministic sampling or cache‐guided prefetching. These results empirically confirm Proposition 1: fixing the PRNG seed and employing a hot‐node cache do not bias or destabilize the stochastic gradient estimates, preserving the convergence guarantees of standard mini-batch SGD.

\section{Conclusion}

We present \name, an access pattern-based cache optimization method and prefetching technique for distributed GNN training. It significantly improves communication overhead and training time without compromising the model's accuracy by actively reducing communication and reusing features. Our implementation requires minimal changes within the DistDGL framework and uses existing modules to build the \name architecture while gaining substantial improvement. On two respective benchmark graphs, we demonstrate significantly better epoch time (reduction in overall training time) and reduction of communication overhead without affecting accuracy. In the future, we also plan to extend this model to other GNN architectures, as our method does not require any modification to existing architecture. We also plan to analyze the performance and energy consumption trade-offs further and design predictive system-level optimizations to increase communication efficiency with minimum memory footprint.

\bibliographystyle{IEEEtran}
\bibliography{references} 

\vspace{12pt}
\color{red}

\end{document}